\documentclass{emulateapj}

\usepackage{graphicx}
\usepackage{subfigure}
\usepackage{amsmath,gensymb}
\usepackage{natbib}
\usepackage{hyperref}
\usepackage[rightcaption]{sidecap}

\newcommand{\kms}{km\,s$^{-1}$}

\newcommand{\om}{\Omega_m}

\newcommand{\oll}{\Omega_\Lambda}
\newcommand{\ob}{\Omega_b}

\newcommand{\beq}{\begin{equation}}
\newcommand{\eeq}{\end{equation}}
\newcommand{\bea}{\begin{eqnarray}}
\newcommand{\eea}{\end{eqnarray}}
\newcommand{\bctr}{\begin{center}}
\newcommand{\ectr}{\end{center}}
\newcommand{\lsim}{\mbox{$\:\stackrel{<}{_{\sim}}\:$} }
\newcommand{\gsim}{\mbox{$\:\stackrel{>}{_{\sim}}\:$} }

\newcommand{\zr}{{\bar z}}
\newcommand{\zp}{z^{\rm pec}}
\newcommand{\vp}{v^{\rm pec}}

\newcommand{\hinvMpc}{$h^{-1}$Mpc}
\newcommand{\zbar}{{\bar z}}

\newcommand{\dLbar}{{\bar d}_{\rm L}}
\newcommand{\dL}{d_{\rm L}}

\newcommand{\Pkzero}{P(k)_{z=0}}

\newcommand{\OM}{\Omega_{\rm M}}

\newcommand{\zpec}{z_{\rm pec}}
\newcommand{\zsunpec}{z_\sun^{\rm pec}}

\newcommand{\zSNpec}{z_{\rm SN}^{\rm pec}}

\newcommand{\sigmazp}{\sigma_z^{\rm pec}}
\newcommand{\sigmazr}{\sigma_{\bar z}}
\newcommand{\sigmaz}{\sigma_z}
\newcommand{\sigmamu}{\sigma_{\mu}}

\newcommand{\chiR}{\tilde{\chi}}
\newcommand{\se}{\sigma_8}

\slugcomment{To appear in The Astrophysical Journal}

\shorttitle{Peculiar velocity effects on SN cosmology}
\shortauthors{Davis et al. 2011}

\begin{document}

\title{The effect of peculiar velocities on supernova cosmology}
%
%

\newcommand{\NUMUQ}{1}       
\newcommand{\NUMDARK}{2}     
\newcommand{\NUMCOLUMBIA}{3}  
\newcommand{\NUMKICP}{4}     
\newcommand{\NUMUCASTRO}{5}  
\newcommand{\NUMFNAL}{6}     
\newcommand{\NUMBOHR}{7}     
\newcommand{\NUMSTOCKASTRO}{8} 
\newcommand{\NUMCAPEMATH}{9} 
\newcommand{\NUMSAAO}{10}     
\newcommand{\NUMAFINST}{11}
\newcommand{\NUMSTOCKPHYS}{12}  
\newcommand{\NUMPORT}{13}     
\newcommand{\NUMRIT}{14}      
\newcommand{\NUMUPENN}{15}    
\newcommand{\NUMPENNSTATE}{16} 
\newcommand{\NUMCAPEASTRO}{17} 

\author{
Tamara~M~Davis,\altaffilmark{\NUMUQ,\NUMDARK}
Lam~Hui,\altaffilmark{\NUMCOLUMBIA}
Joshua~A.~Frieman,\altaffilmark{\NUMKICP,\NUMUCASTRO,\NUMFNAL}
Troels~Haugb{\o}lle,\altaffilmark{\NUMBOHR}
Richard~Kessler,\altaffilmark{\NUMUCASTRO,\NUMKICP}
Benjamin~Sinclair,\altaffilmark{\NUMUQ}
Jesper~Sollerman,\altaffilmark{\NUMBOHR,\NUMSTOCKASTRO}
%
%
Bruce~Bassett,\altaffilmark{\NUMCAPEMATH,\NUMSAAO,\NUMAFINST}
John~Marriner,\altaffilmark{\NUMFNAL}
Edvard~M\"ortsell,\altaffilmark{\NUMSTOCKPHYS}
Robert~C.~Nichol,\altaffilmark{\NUMPORT}
Michael~W.~Richmond,\altaffilmark{\NUMRIT}
Masao Sako,\altaffilmark{\NUMUPENN}
Donald~P.~Schneider,\altaffilmark{\NUMPENNSTATE}
Mathew~Smith\altaffilmark{\NUMCAPEASTRO}
%
%
%
%
}  


\altaffiltext{\NUMUQ}{
School of Mathematics and Physics, University of Queensland, QLD,  
4072, Australia
}
\altaffiltext{\NUMDARK}{
Dark Cosmology Centre, Niels Bohr Institute, 
University of Copenhagen, Juliane
Maries Vej 30, DK-2100 Copenhagen \O, Denmark
}
\altaffiltext{\NUMCOLUMBIA}{
Department of Physics, Institute for Strings, Cosmology and Astroparticle Physics (ISCAP),
Columbia University, New York, NY 10027
}
\altaffiltext{\NUMKICP}{
Kavli Institute for Cosmological Physics, 
The University of Chicago, 5640 South Ellis Avenue Chicago, IL 60637
}
\altaffiltext{\NUMUCASTRO}{
  Department of Astronomy and Astrophysics,
   The University of Chicago, 5640 South Ellis Avenue, Chicago, IL 60637
}
\altaffiltext{\NUMFNAL}{
Center for Particle Astrophysics, 
  Fermi National Accelerator Laboratory, P.O. Box 500, Batavia, IL 60510
}
\altaffiltext{\NUMBOHR}{
Niels Bohr Institute, 
University of Copenhagen, Juliane
Maries Vej 30, DK-2100 Copenhagen \O, Denmark
}
\altaffiltext{\NUMSTOCKASTRO}{
The Oskar Klein Centre, Department of Astronomy, AlbaNova, 
Stockholm University, SE-106 91 Stockholm, Sweden
}
\altaffiltext{\NUMCAPEMATH}{
Department of Mathematics and Applied Mathematics,
University of Cape Town, Rondebosch 7701, South Africa
}
\altaffiltext{\NUMSAAO}{
  South African Astronomical Observatory,
   P.O. Box 9, Observatory 7935, South Africa
}
\altaffiltext{\NUMAFINST}{
African Institute for Mathematical Sciences,
6 Melrose Road, Muizenberg, Cape Town,
 South Africa
}
\altaffiltext{\NUMSTOCKPHYS}{
Department of Physics, AlbaNova, Stockholm University, 
SE-106 91, Stockholm, Sweden
}
\altaffiltext{\NUMPORT}{
Institute of Cosmology and Gravitation, Dennis Sciama Building,
Burnaby Road, University of Portsmouth, Portsmouth, PO1 3FX, UK
}
\altaffiltext{\NUMRIT}{
  Physics Department,
   Rochester Institute of Technology,
   85 Lomb Memorial Drive, Rochester, NY 14623-5603
}

\altaffiltext{\NUMUPENN}{
Department of Physics and Astronomy,
University of Pennsylvania, Philadelphia, PA  19104
}

\altaffiltext{\NUMPENNSTATE}{
  Department of Astronomy and Astrophysics,
   The Pennsylvania State University,
   525 Davey Laboratory, University Park, PA 16802
}
\altaffiltext{\NUMCAPEASTRO}{
 Astrophysics, Cosmology and Gravity Centre, 
 Department of Mathematics and Applied Mathematics,
 University of Cape Town, Rondebosch 7701, South Africa
 }

\email{tamarad@physics.uq.edu.au}

\begin{abstract}

We analyse the effect that peculiar velocities have on the cosmological inferences we make using luminosity distance indicators, such as type Ia supernovae.  In particular we study the corrections required to account for
(a) our own motion, 
(b) correlations in galaxy motions, and
(c) a possible local under- or over-density.
For all of these effects we present a case study showing the impact on the cosmology derived by the Sloan Digital Sky Survey-II Supernova Survey (SDSS-II SN Survey). 

Correcting supernova redshifts for the cosmic microwave background (CMB) dipole slightly over-corrects nearby supernovae that share some of our local motion.  We show that while neglecting the CMB dipole would cause a shift in the derived equation of state  of $\Delta w\sim0.04$ (at fixed $\om$) the additional local-motion correction is currently negligible ($\Delta w\lsim0.01$).

We then demonstrate a covariance-matrix approach to statistically account for correlated peculiar velocities.  This down-weights nearby supernovae and effectively acts as a graduated version of the usual sharp low-redshift cut.  Neglecting coherent velocities in the current sample causes a systematic shift of $\Delta w\sim0.02$.  This will therefore have to be considered carefully when future surveys aim for percent-level accuracy and we recommend our statistical approach to down-weighting peculiar velocities as a more robust option than a sharp low-redshift cut.

\end{abstract}



\section{Introduction}\label{intro}

Concordance cosmology ($\Lambda$CDM) is a successful model of our universe, fitting observations of type Ia supernovae \citep{riess98,perlmutter99,astier06,WV07,riess04,riess07,kowalski08,hicken09,kessler09,freedman09}, the cosmic microwave background \citep{page03,tegmark06,spergel07,komatsu09,komatsu11}, baryon acoustic oscillations \citep{eisenstein05,percival07,percival10,blake11b}, and growth of structure \citep{blake11a}, amongst others.  However, it relies on the existence of dark components of the universe -- dark energy and dark matter -- whose nature remains mysterious.  This has given rise to questions about the validity of our theory of gravity itself.  Therefore enormous observational effort is continuing to better characterise the dark sector by measuring the expansion history of the universe and the growth of structure within it.  

Type Ia supernovae remain a lynch-pin in this effort, and more surveys are underway, or planned, to gather ever more high quality data to try to reduce the uncertainties on our cosmological parameters down below the 1\% level and search for possible variations in the equation of state of dark energy.   To achieve this accuracy we will have to address small systematic effects that had previously been negligible.  In this paper we consider systematic errors that could arise from neglecting the peculiar velocities and gravitational redshifts induced by large scale structure.  

The customary diagnostic in supernova cosmology is the Hubble diagram, a measurement of luminosity as a function of redshift.  %
When using type Ia supernovae to measure this magnitude-redshift relation, the redshift used should be entirely due to the expansion of the universe.  In practice this is never the case, as large-scale structure in the universe induces peculiar motions so that the measured redshift contains some contribution from peculiar velocities.  To date, the majority of the effort in calibrating type Ia supernova measurements has been increasing the accuracy and precision with which we can determine their luminosity, and thus their use as a standard candle.  In comparison the uncertainty on the redshift of the supernovae has usually been considered negligible.  It is this more neglected uncertainty we turn our attention to in this study. 

We distinguish between the cosmological redshift, $\bar{z}$ due entirely due to the expansion of the universe; the `peculiar' redshift, $\zpec$, due entirely to peculiar velocities; and the gravitational redshift, $z_{\rm grav}$, due to density fluctuations.  These three redshifts combine to give the observed redshift, $z$, according to,
\beq (1+z) = (1+\bar{z})(1+\zpec)(1+z_{\rm grav}).\eeq

Systematic peculiar velocity effects can manifest themselves in a number of ways.  In analogy to the cosmic microwave background (CMB), we can consider monopole, dipole, and higher-multipole effects.  
\begin{itemize}
\item Monopole effects mimic expansion by imprinting a non-cosmological redshift isotropically on all sources.  They can arise if we are in the center of a local over- (or under-)density due to both peculiar velocities of infall (or outfall) and gravitational redshifts.  Due to their isotropic nature they are very difficult to distinguish from cosmological redshifts and could lead to significant systematic errors in cosmological parameter estimation.
\item Dipole effects occur due to our own peculiar motion, or due to being off-center in a local over- or under-density.  Our motion with respect to the overall expansion is well measured by the dipole in the CMB, and can be easily corrected-for.  However, the correction needs to be modified when calculating our velocity relative to local galaxies that share some of our motion with respect to the CMB.
\item Higher-order multipole effects occur when distant sources move coherently as they share gravitational attraction to common large-scale structure.  These are analogous to the fluctuations seen in the CMB after the monopole and dipole terms have been removed.  
\end{itemize}
All these peculiar velocities are manifestations of coherent flows in the universe.  We consider the lowest order coherent flows (monopole and dipole) separately because they have the largest systematic effects.  

\begin{figure}
\plotone{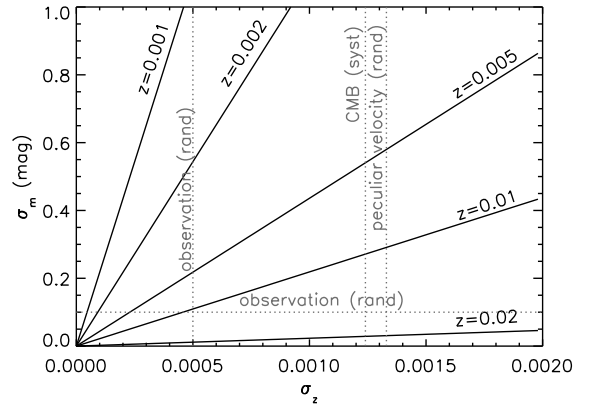}
\caption{Plot of the conversion between a redshift uncertainty $\sigma_z$ and a magnitude uncertainty $\sigma_m$, for a variety of redshifts between $z=0.001$ and $z=0.02$ as labelled.  Dotted lines depict the typical redshift dispersion from random peculiar velocities, and  the observational uncertainties in magnitude and redshift (the observational uncertainty for $\sigma_z$ assumes we have redshifts from host galaxy spectra, when only SN spectra are available the redshift error is an order of magnitude higher).   Those random contributions are all marked `(rand)', while the other dotted line depicting `CMB (syst)' shows the maximum systematic shift in redshift caused by the CMB dipole.  
This figure shows why the low-redshift cutoff of $z=0.02$ is appropriate, because above this redshift all redshift uncertainties are significantly smaller than the observational magnitude uncertainty.  The conversion has been done for the $(\om,\oll)=(0.3,0.7)$ model, but there is negligible difference between this and most plausible homogeneous cosmological models over this redshift range (including the empty model and a model with $\om=0.3$ but no cosmological constant).  }
\label{fig:dmdz}
\end{figure}

When we compare the precision of redshift ($z$) measurements with the precision of apparent magnitude ($m$) measurements we are primarily concerned with their effect on the Hubble diagram (rather than the relative precision as measured by $\Delta z/z$ and $\Delta m/m$).  The effect on the Hubble diagram can be quantified by considering the slope of the magnitude-redshift relation, $dm/dz$.  Indeed, the standard method for including redshift uncertainties ($\sigma_z$) in cosmological analyses is to convert them to magnitude uncertainties ($\sigma_m$) using the $dm/dz$ derived from a fiducial cosmological model (see Appendix~\ref{app:dmdz} for an outline of the procedure). The conversion between a redshift uncertainty and a magnitude uncertainty is shown in Figure~\ref{fig:dmdz}, for a few different redshifts.  At higher redshifts the magnitude-redshift relation is flatter ($dm/dz$ is smaller) which means large redshift uncertainties generate only small magnitude uncertainties.  At low redshifts the converse is true, and small redshift uncertainties give large magnitude uncertainties.  

The measurement of redshift will remain far more accurate than the measurement of the supernova magnitude into the foreseeable future.  However, the accuracy of those measurements can be misleading since systematic effects on redshift due to peculiar motions can be much larger than the measurement error (see Fig.~\ref{fig:dmdz}). 

The dominant source of intrinsic redshift dispersion (that is always included in cosmological analyses) is the effect of random peculiar velocities.  These are usually taken to be about\footnote{For the SDSS sample \citet{kessler09}  use $\sigma_v^{\rm pec}=300$\kms\, for the random peculiar motions added in quadrature with $\sigma_v^{\rm pec}=200$\kms\, for the internal velocities, giving a total $\sigma_v^{\rm pec}$=360\kms, corresponding to $\sigma_z^{\rm pec}$ of 0.0012.  For the ESSENCE sample \citet{WV07} uses $\sigma_v^{\rm pec}=400$\kms.} 
$\sigma_v^{\rm pec}=300$\kms, which according to $v^{\rm pec}=cz^{\rm pec}$ corresponds to an error in redshift of $\sigma_z^{\rm pec} = 0.001$.   This redshift uncertainty gives a non-negligible magnitude uncertainty of $\sigma_m^{\rm pec}  = 0.2$ for objects at redshift $z=0.01$, which reduces to  $\sigma_m^{\rm pec} = 0.02$ for objects at $z=0.1$.  These values should be compared with the intrinsic diversity in supernova magnitudes of $\sigma_m^{\rm int}\approx 0.1$ and the observational magnitude uncertainty (including the uncertainty in fitting the SN light curves) of $\sigma_m^{\rm meas}\lsim 0.1$ for the most distant supernovae included in most samples (closer ones have smaller observational uncertainty).  

The observational uncertainty in redshift is typically $\sigma_{z}^{\rm spec}\approx$ 0.0005 for SN host-galaxy-based redshifts measured by SDSS, and $\sigma_{z}^{\rm spec}\approx$ 0.005 for redshifts based on the supernova spectra alone \citep{zheng08}.  So the observational uncertainty is of similar magnitude to the intrinsic scatter due to peculiar motions. 

Treating these motions as random scatter is not entirely valid, since galaxies (and the supernovae in them) preferentially fall into overdense regions, so objects in the same region of sky will tend to have correlated peculiar velocities.   Detailed measurements of this effect may provide another technique for measuring the matter distribution of the universe and thus derive cosmological parameters using diagnostics such as the peculiar velocity power spectrum \citep[e.g.][]{bonvin06flucts,neill07,gordon07,gordon08,abate08,hannestad08}.   \citet[][App.~B]{lampeitl10} showed that the signal in the SDSS SN data set is too small to measure cosmological parameters this way (mostly because our sample is too distant).  Here we therefore concentrate only on the deleterious impact peculiar velocities have on the cosmological results derived from the SN magnitude-redshift relation.

\subsection{Data}
In this paper we use the Sloan Digitial Sky Survey-II  Supernova Survey \citep[SDSS-II SN Survey][]{york00,holtzman08,frieman08,sako08} as a case study.  For the SDSS-II SN Survey repeat images were taken of an equatorial stripe, 2.5 degrees wide and about 120 degrees long centered on RA~23.5hr (SDSS Stripe 82), 
 which corresponds to a field center at Galactic coordinates $(\ell,b)\approx(84\degree,-57\degree)$.  The direction of the CMB dipole in Galactic coordinates is toward $(\ell,b)=(263\degree.85\pm0\degree.1,48\degree.25\pm0\degree.04)$ \citep{bennett03};  so the antipode direction lies at $(\ell,b)\approx(83\degree.85,-48\degree.25)$.   That means that the center of the SDSS field is almost aligned with the direction in which the CMB appears blue-shifted.

Over three years about 500 spectroscopically confirmed Type Ia supernovae were discovered in the redshift range $0.05<z<0.4$.  The first year's data including 103 supernovae were published in \citet{holtzman08} and analysed in \citet[][hereafter referred to as K09]{kessler09}\footnote{\url{http://das.sdss.org/va/SNcosmology/sncosm09_fits.tar.gz}} who combined these with a re-analysis of existing data to make a coherent sample of 288 supernovae that were used to measure cosmological parameters.  It is this data set that we use here, focussing on several of the subsets they define to demonstrate the effect of different redshift ranges (see Table~\ref{tab:data}).  

Since the SDSS sample concentrates on relatively nearby supernovae ($0.05\lsim z \lsim 0.4$) the peculiar velocity contribution is a more significant proportion of the total redshift than in surveys that focus on higher redshifts such as  ESSENCE \citep{WV07}, SuperNova Legacy Survey \citep{astier06}, and Higher-$z$ \citep{riess07}.  Moreover, since the region of sky covered by the SDSS SN survey lies close to the direction of the CMB dipole, (Fig.~\ref{fig:map}) the alignment conspires to maximise the magnitude of the effect.  For all these reasons the SDSS SN sample is an interesting one in which to test peculiar velocity effects. 

\begin{deluxetable}{cccccc}
\tablecolumns{6}
\tablewidth{0pc}
\tablecaption{Sub-samples of type Ia supernovae used by Kessler et al.\ 2009.  The top three rows give the survey name, redshift range, and number of supernovae in each data set.  Below the line, ticks indicate which data sets are included in each of the sub-samples.}
\tablehead{
\colhead{} & \colhead{Nearby}   & \colhead{SDSS-II}    & \colhead{ESSENCE} &
\colhead{SNLS}    & \colhead{HST}   \\
\colhead{$z$:} & \colhead{0.02--0.10} & \colhead{0.04--0.42} & \colhead{0.16--0.69} & \colhead{0.25--1.01} & \colhead{0.21--1.55} \\
\colhead{$N$:} & \colhead{33} & \colhead{103} & \colhead{56} & \colhead{62} & \colhead{34} 
}
\startdata
a: &  & \checkmark & & & \\ 
b: & & \checkmark & \checkmark & \checkmark & \\ 
c: & \checkmark & \checkmark & & & \\ 
d: & \checkmark & \checkmark & \checkmark & \checkmark & \\ 
e: & \checkmark & \checkmark & \checkmark & \checkmark & \checkmark \\ 
f: & \checkmark & & \checkmark & \checkmark & 
\enddata
\label{tab:data}
\end{deluxetable}

\subsection{Notation}
We use the metric 
\beq ds^2 = -c^2 dt^2 + R(t)^2\left[d\chi^2 + S_k^2(\chi)d{\mathbf \Theta^2}\right],\eeq
where $t$ is proper time, $R(t)$ is the scale factor with dimensions of length, $\chi$ is the dimensionless comoving coordinate, $S_k(\chi)=\sin(\chi), \chi, \sinh(\chi)$ for closed, flat, and open universes respectively, and $d{\mathbf \Theta}$ encompasses the angular terms.   The present day scale factor, $R_0 \equiv c/(H_0\sqrt{|\Omega_k|})$, and the dimensionless scale factor is defined as $a=R/R_0$.  Hubble's parameter is $H(z)=\dot{a}/a$, where an overdot represents differentiation with respect to proper time.  The dimensionless comoving distance as a function of redshift is related to cosmological parameters by
\beq \chi = \frac{c}{R_0}\int_0^z\frac{dz}{H(z)}.\eeq
Frequently we will need the comoving distance with units of length, for which we will use the shorthand $\chiR \equiv R_0\chi$.

We begin in Section~\ref{sect:theory} by providing the theory needed to make dipole and correlated velocity corrections to distance measurements.  Although focusing on type Ia supernova measurements this analysis is general to any luminosity distance measure.  In Section~\ref{sect:impact} we then perform a case study of these effects on the cosmological inferences made using the supernova data from the SDSS II supernova survey \citep{kessler09}.  We conclude in Section~\ref{sect:conclusion}.

\section{Correcting for peculiar velocities}\label{sect:theory}
\subsection{Dipole correction}\label{sect:dipole}

When using supernova redshifts to make cosmological inferences we need to remove the imprint of our own peculiar motion so that the redshift of the supernova is entirely due to the expansion of the universe.  To first order this is straightforward, since we know our own velocity to high precision from measurements of the CMB dipole.  Correcting for the CMB dipole is standard practise in all supernova cosmology analyses \citep[e.g.][]{astier06,WV07,riess07,kowalski08,kessler09}.  However, no correction is typically made for our motion relative to nearby galaxies, for which the CMB dipole is a poor approximation.  Here we investigate the impact of both effects. 


We are moving at ${\mathbf v}_\sun^{\rm pec}=371$~\kms\, with respect to the CMB  \citep{kogut93,bennett03}. 
This is small compared to the Hubble flow for all but the nearest objects, dropping to less than $\sim 1$\% beyond a redshift of 0.1.   Our motion thus contributes a maximum redshift change of $\sigma_z=0.00124$ to sources that are directly aligned with the dipole.  This is much less than the equivalent uncertainty in our magnitude measurement  (see Fig.~\ref{fig:dmdz}); it is only its coherence among SNe that can make it significant.


The general relationship between redshift and {\em peculiar} velocity is, 
\beq 1+\zp = \sqrt{\frac{1+\vp/c}{1-\vp/c}},\label{eq:zp}\eeq
which simplifies to $\zp=\vp/c$ in the non-relativistic limit.\footnote{Note that the special relativistic velocity-redshift relation in Eq.~\ref{eq:zp} is only appropriate for peculiar velocities.  It is not appropriate for recession velocities (the velocity that appears in Hubble's Law), for which special relativistic corrections should {\em never} be applied \citep{davis04,davis05}.}   
The redshift correction required to account for our velocity with respect to the CMB, ${\mathbf v}_\sun$, is
\beq z_\sun^{\rm pec} = -v_\sun^{\rm pec}/c = {\mathbf v}_\sun^{\rm pec} \cdot (\mathbf{-n})/c \eeq
where ${\mathbf n}$ is the unit vector from the sun to the supernova.  
(The negative sign ensures that if we are moving in the direction of the supernova the resulting correction is a blueshift.)

The observed heliocentric redshift, $z$, is then related to the cosmological redshift, $\zbar$, by,\footnote{This assumes the observer has already corrected both for the motion of the Earth around the Sun, which contributes up to 30\kms\, depending on the time of observation 
and for atmospheric refraction, which contributes up to 90\kms\, (the index of refraction of air is 1.0003, so $\Delta z=0.0003$ and $c\Delta z=90$\kms).  Usually this is done as a standard step in wavelength calibration.}
\beq (1+z) = (1+\zbar)(1+z_\sun^{\rm pec}). \label{eq:zgood}\eeq
Note that the NED velocity calculator \citep{NED} uses the approximation, 
\beq  \quad z \approx \zbar + z_\sun^{\rm pec} .\label{eq:zapprox}\eeq
This gives a fractional error\footnote{Rearranging Eq.~\ref{eq:zgood} and Eq.~\ref{eq:zapprox} gives $\frac{\zbar^{\rm NED}-\zbar}{\zbar} = \zsunpec$.} of precisely $\zsunpec$, which is negligible for most circumstances.

The dipole not only shifts the redshift but also changes the apparent magnitude of the source due to the Doppler shift of the photon energy and relativistic beaming.  
The CMB dipole therefore also has an effect on the {\em luminosity distance} calculated from the magnitude of a supernova \citep{sasaki87,pyne96,bonvin06flucts,cooray06,hui06}.   
This arises because the luminosity distance is related to the comoving distance, $\chi$, by (recalling that overbars refer to observations made from the CMB rest frame),
\beq \dLbar(\zbar) =  (1+\zbar)R_0S_k(\chi). \eeq
However, what we actually observe is (recalling that $z$ is the observed redshift and considering for the moment only our own motion),
\bea \dL(z) &=& (1+z)R_0S_k(\chi), \label{eq:dlz}\\
&=& (1+\zbar)(1+z_\sun^{\rm pec})R_0S_k(\chi), \label{eq:dlzbar}\\
&=& \dLbar(\zbar)(1+z_\sun^{\rm pec}).\eea
So both the redshift and the luminosity distance need to be corrected for the effect of the dipole.\footnote{You may be concerned that in going from Eq.~\ref{eq:dlz} to Eq.~\ref{eq:dlzbar} we've neglected the factor of $z$ in the calculation of $\chi=(c/R_0)\int_0^\zbar d\zbar/H(\zbar)$.  However, this cosmological redshifting is independent of the motion of the emitter or observer, and therefore does not need correcting for peculiar velocities.  As long as we correct the redshift of the supernova to the CMB frame our theoretical model comparison will be correct.}  

Alternatively, one can choose to correct for both in one fell swoop by correcting the observed luminosity distance at redshift $z$ to the luminosity distance that would have been observed at redshift $z$ in the absence of peculiar velocities.  This is the approach  
taken by the commonly used program simple\_cosfitter \citep{conley06}.\footnote{\href{http://qold.astro.utoronto.ca/conley/simple\_cosfitter/html/}{http://qold.astro.utoronto.ca/conley/simple\_cosfitter/html/}}   \citet{hui06} give the formula for $\dLbar(z)$, which 
can be used to correct only the $d_L$ values without correcting $z$.   Considering only our own motion, Eq.~15 of \citet{hui06} can be rearranged to give,
\beq \dL(z) = \dLbar(z)\left[ 1+ \frac{a_e}{a'_eR_0T_k(\chi)}\mathbf{v_{0}.n}\right], \label{eq:dlz2} \eeq
where $a'_e \equiv da_e/d\tau$ represents the derivative of the scale factor with respect to conformal time, evaluated at the time of emission,\footnote{
We give conformal time dimensions of time, so $d\tau = dt/a$ and the conformal time derivative is related to the proper time derivative (denoted by an overdot) according to
\beq a' = \frac{da}{dt}\frac{dt}{d\tau} = \dot{a} a.\eeq
} and here we have kept the curvature dependence explicit, with $T_k(\chi)\equiv\tan(\chi), \chi$, and $\tanh(\chi)$ in closed, flat, and open universes respectively.


\begin{figure}
\plotone{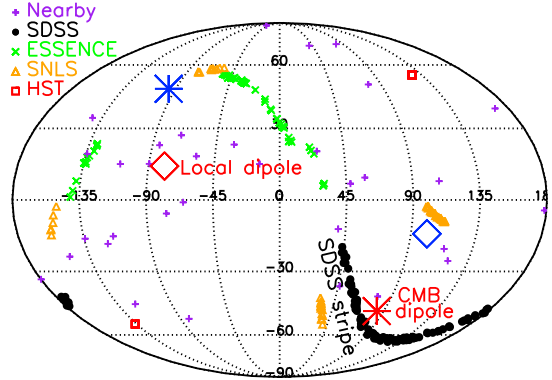}
\caption{Map of the distribution of type Ia supernovae in the Nearby, SDSS, ESSENCE, SNLS, and HST samples in Galactic coordinates with the local ($z\sim0.1$) dipole indicated by diamonds and the CMB dipole indicated by stars.   The center of the SDSS-II SN stripe lies close to the direction of the CMB dipole, which makes it important to carefully correct the SDSS sample for the effects of the dipole.
The local dipole is measured with respect to the CMB and is in approximately the opposite direction since galaxies in our local neighbourhood tend to share some of our peculiar velocity with respect to the CMB. }
\label{fig:map}
\end{figure}

When there are two peculiar velocities to correct, such as when accounting for the supernova's motion\footnote{The additional redshift due to the supernova's motion is 
\beq z_{\rm SN}^{\rm pec} = v_{\rm SN}^{\rm pec}/c = {\mathbf v}_{\rm SN}^{\rm pec} \cdot \mathbf{n}/c ,\eeq
where again ${\mathbf n}$ is the unit vector from the sun to the supernova and ${\mathbf v}_{\rm SN}^{\rm pec}$ is measured with respect to the CMB.}
 with respect to the CMB ($z_{\rm SN}^{\rm pec}$) in addition to our own motion, one uses
\beq (1+z) = (1+\zbar)(1+z_\sun^{\rm pec})(1+z_{\rm SN}^{\rm pec}).\label{eq:zzz}\eeq
This equation is valid even for relativistic velocities, but in the literature it is more common to encounter approximate formulae such as \citep{hui06}, 
\bea (1+z) &=& (1+\zbar)(1 - {\mathbf v}_\sun \cdot {\mathbf n}/c +{\mathbf v}_{\rm SN} \cdot {\mathbf n}/c), \\
 &=&(1+\zbar)(1+z_\sun^{\rm pec}+z_{\rm SN}^{\rm pec}),\eea
which are perfectly appropriate for the low velocities we encounter in almost all practical situations. 

When including the source motion the correction to the luminosity distance becomes,
\beq \dL(z) = \dLbar(\zbar)(1+z_\sun^{\rm pec})(1+z_{\rm SN}^{\rm pec})^2.\eeq
Note that two factors of $(1+\zSNpec)$ enter the luminosity distance correction.  One is due to the Doppler shifting of the photons, the other is due to relativistic beaming.


\begin{figure}
\plotone{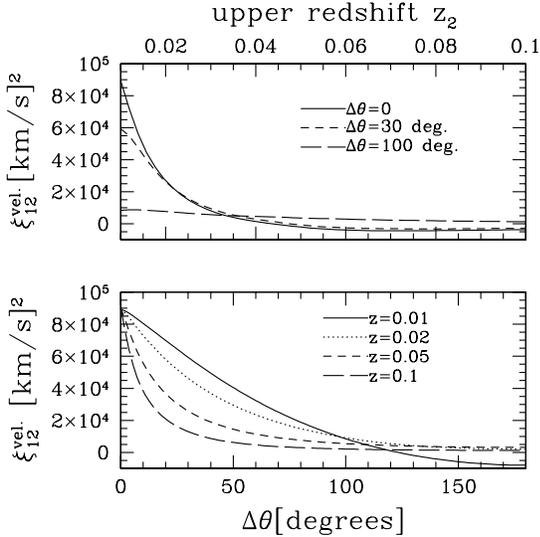}
\caption{\footnotesize The velocity two-point correlation, $\xi_{12}$, in units of $($\kms$)^2$.  The upper panel shows the correlation as a function of $z_2$ (the redshift of the higher redshift supernova) compared to a low-redshift supernova at a fixed $z_1=0.01$, for three different angular separations (0, 30 and 100 degrees).  As the redshift separation increases the correlation diminishes.  The lower panel shows the correlation as a function of angular separation, in the case where the two supernovae are at the same redshift ($z_2=z_1=$0.01, 0.02, 0.05, and 0.1).  For a fixed angular separation the correlation is most dramatic at low redshifts because this corresponds to a smaller physical distance than the same angular separation at high redshifts. $\xi_{12}$ is given by the last two lines of Eq.~\ref{cijvellarge}. \vspace{5mm}}
\label{fig:xiV}
\end{figure}

\begin{figure}
\plotone{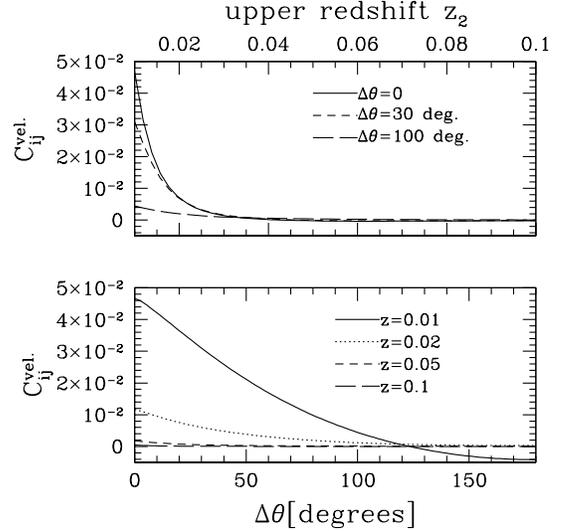}
\caption{\footnotesize The magnitude covariance due to peculiar motion (Eq.~\ref{cijvellarge}).  As for Fig.~\ref{fig:xiV} the upper panel shows, for three different angular separations, the covariance between a $z_1=0.01$ source and a source at a higher redshift, $z_2$.  The lower panel shows for fixed redshifts ($z_2=z_1=$0.01, 0.02, 0.05, and 0.1) how the covariance drops as the angular separation increases.  To get a feel for how much additional uncertainty the correlated component adds to the uncorrelated dispersion consider that for these four redshifts the canonical 300\kms\ dispersion corresponds to $\sigma_m^2\sim(25, 10, 7, 5)\times10^{-3}$ respectively (see Appendix~\ref{app:dmdz}).  So the strongest correlated uncertainties are an order of magnitude lower than the random dispersion.  However, the random dispersion can be beaten down by raising the number of SNe, while the correlated covariance cannot.  Thus, as the total number of SNe increases the correlated noise becomes comparatively more and more important.    \vspace{5mm}}
\label{fig:Cv}
\end{figure}


\subsection{Coherent flows}\label{sect:coherent}\label{sect:correlated}

The next peculiar velocity effect we consider is the impact of coherent bulk motions. 
Large scale structure introduces correlated peculiar velocities as neighbouring galaxies, and the supernovae they host, fall towards the same overdensities.  Ignoring these correlations underestimates the uncertainty in our cosmological inferences.  The effect is particularly important at low redshift where supernovae will tend to be physically closer to each other (as a function of angular separation on the sky). 

These coherent flows are the higher-order manifestations of the gravitational influence of large scale structure, beyond the local dipole we discussed in Sect.~\ref{sect:dipole}.  This exercise is analogous to subtracting the dipole from the CMB and considering the residual fluctuations. 

The reason correlated velocities have a deleterious impact on standard supernova cosmology arises because two correlated supernovae can not be statistically averaged in such a way so as to reduce the error in proportion to the square-root of the number of supernovae, as is usually assumed.  Correlations mean the data are not randomly distributed about the central value, so the uncertainties do not average out exactly and there remains a residual error even in the limit of an infinite number of data points.

Here we calculate the expected statistical correlations in the peculiar velocities of galaxies as a function of their distance from each other based on a $\Lambda$CDM model.  We convert this into an observationally useful measure by converting it to the expected covariance in magnitudes as a function of angular separation and redshift separation.

Recall that supernova cosmology aims to fit the observations of apparent magnitude, $m$, and redshift, $z$ of a supernova to the following relation,
\beq m(z) = 5 {\,\rm log}_{10} d_L(z) + {\cal M} ,\eeq
where $d_L$ is the luminosity distance (in units of 10pc) and ${\cal M}$ is a constant incorporating the absolute magnitude of the supernova and Hubble's constant.  The luminosity distance is a function of the cosmological parameters we want to fit.

The likelihood of a particular model, in a gaussian distribution, is proportional to $e^{-{\mathbf \chi}^2/2}$.
Let $\hat{m}_i$ represent the $i$th measurement and $m_i$ the corresponding model prediction. When all data points are independent, $\chi^2$ is given by,\footnote{In this paragraph $\chi^2$ represents the statistic, not the comoving coordinate squared.} 
\beq \chi^2 = \sum_{i}\frac{(\hat{m}_i-m_i)^2}{\sigma_i^2}.\eeq
Here $\sigma_i \equiv \sigma_{m_i}$ is the magnitude uncertainty on $m_i$.
However, when data points are correlated the more general form of $\chi^2$ is given by,
\beq \chi^2 = \sum_{i,j} (\hat m_i - m_i) C^{-1}_{ij} (\hat m_j - m_j),\eeq 
where the covariance matrix,
\beq C_{ij} \equiv \langle \delta\hat{m}_i\delta\hat{m}_j\rangle, \eeq
quantifies how likely two supernovae are to have the same offset from the correct model.   The factor $\delta \hat{m}_i \equiv \hat{m}_i-\langle\hat{m}_i \rangle$ designates how far the $i$th data point deviates from the mean of the observational data.  Correlations in these deviations arise for several reasons in addition to the peculiar velocity affect we consider here, for example \citet{blondin10} studies the covariance between distance prediction errors due to magnitude error, peculiar velocities, and intrinsic covariance. 

The covariance matrix can be divided into the random component $\sigma_i$ and the correlated component, which we consider to be only due to peculiar velocities, $C_{ij}^{\rm vel}$,  
\beq C_{ij} = \sigma_i^2\delta_{ij} + C_{ij}^{\rm vel}.\label{Cfull}\eeq
As discussed in \citet{hui06}, there are a wider variety of large scale structure induced fluctuations
than are accounted for in Eq.~\ref{Cfull}. For instance, lensing introduces correlated noise
in addition to Poissonian fluctuations. There are also fluctuations due to
gravitational redshift and the integrated Sachs-Wolfe effect. It can be shown that all these effects
can be neglected for surveys of current practical interest \citep{hui06}.

The random uncertainties, which contribute to the diagonal part of $C_{ij}$, include the intrinsic diversity in the supernova population, $\sigma_i^{\rm intr}$, the scatter due to measurement uncertainty, $\sigma_i^{\rm meas}$, and the contribution from random peculiar velocities, $\sigma_i^{\rm vel}$.  Although random peculiar velocities primarily add dispersion in the apparent redshift of the sources (the effect on the luminosity is smaller and is usually neglected), this is usually converted to a dispersion in magnitude and added in quadrature to the other magnitude uncertainties (see Appendix~\ref{sect:random} for more detail),
\beq \sigma_i^2 = {\sigma_i^{\rm intr}}^2+{\sigma_i^{\rm meas} }^2+{\sigma_i^{\rm vel}}^2.\eeq 
Since random peculiar velocities are taken into account by this diagonal term we set all diagonal terms $C_{ij}^{\rm vel}=0$.  Alternatively we could remove $\sigma_i^{\rm vel}$ and reinstate them as the diagonal elements of  $C_{ij}^{\rm vel}$.  

The velocity correlation function is defined to be, 
\beq \xi^{\rm vel}_{ij} \equiv \langle ({\bf v_i} \cdot {\bf \hat x_i}) ({\bf v_j} \cdot {\bf \hat x_j})
\rangle \eeq
where ${\bf \hat x_i}$ and ${\bf \hat x_j}$ represent the unit vectors pointing towards SNe $i$ and $j$ respectively, and ${\bf v_i}$ and ${\bf v_j}$ represent the velocity vectors of each supernova's motion.

The peculiar-motion-induced magnitude covariance is related
to the velocity correlation function $\xi_{ij}^{\rm vel}$ 
by,
\begin{eqnarray}
C_{ij}^{\rm vel} = \left[{5 \over {c\,\rm ln\,} 10}\right]^2
\left[1 - {a_i \over a'_i} {c \over \chiR_i} \right] 
\left[1 - {a_j \over a'_j} {c \over \chiR_j}\right] \xi^{\rm vel}_{ij} , \\ \nonumber
\end{eqnarray}
where $c$ is the speed of light, $\chiR\equiv R_0\chi$ is the radial comoving distance, $a=R/R_0$ is the normalised scale factor,
and the prime denotes the conformal time derivative.  All quantities
with a subscript $i$ or $j$ are to be evaluated at the redshift of the SN in question.  For a non-flat universe $\chiR\rightarrow R_0T_k(\chi)$.

A numerical code to compute both $\xi_{ij}^{\rm vel}$ and $C_{ij}^{\rm vel}$ 
for a pair of points at arbitrary redshifts and angular separation in the standard cosmological model of $\Lambda$CDM
is available at {\tt http://www.astro.columbia.edu/$\sim$lhui/PairV}.  We illustrate the results of that code in Figs.~\ref{fig:xiV} and \ref{fig:Cv}, and in what follows we explain the theory behind those calculations. 

To calculate the expected velocity correlation function given a theoretical model we need information about how structure grows.  To first order this is given by the linear growth factor $D(z)\equiv\delta(z)/\delta(0)$, where the overdensity $\delta = (\rho-\langle\rho\rangle)/\langle\rho\rangle$.  
As input we also use the mass power spectrum of density fluctuations observed at the present time $\Pkzero$, where $k$ is the comoving wavenumber 
(inverse distance).  

Using these we can estimate the distribution of peculiar velocities expected in a particular theoretical model.  Concentrating for the moment only on the dispersion (the diagonal terms in the velocity correlation function), one finds the dispersion in peculiar velocities to be \citep{hui06},
\beq {\sigma^{\rm vel}_{v_i}}^2 \equiv \xi^{\rm vel}_{ii} = {D'(z_i)}^2 \int_0^\infty {dk \over 6 \pi^2} \Pkzero ,\eeq
which results in a dispersion in apparent magnitude of,
\beq
\label{sigPfull}
{\sigma_i^{\rm vel}}^2 =
\left[\frac{5}{c \ln 10}\right]^2 
\left[1 - {a_i \over a'_i} {c \over \chiR_i}\right]^2 {\sigma^{\rm vel}_{v_i}}^2. 
\eeq
In principle this dispersion is sensitive to non-linear fluctuations, 
but the velocity power spectrum weights larger-scale modes more than the density power spectrum does and we find that when using the linear mass power spectrum for a $\Lambda$CDM model 
the resulting value for $\sigma_{v_i}$ 
agrees with the canonical value of $300$ km/s to better than
$10 \%$ for all redshifts of interest.  
The off-diagonal components of $C_{ij}$ should be at least as well fit by linear theory since they are less sensitive to small-scale structure than $\sigma_{v_i}$. 

\begin{figure}
\plotone{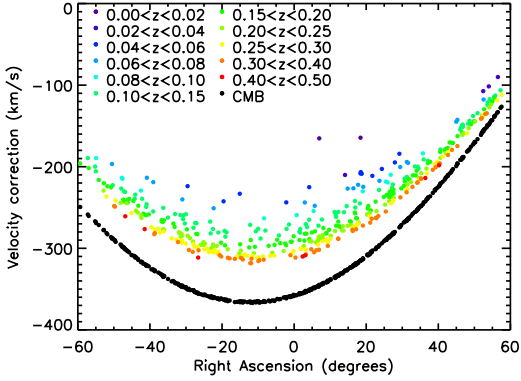}
\caption{Peculiar velocity correction required for the SDSS sample, plotted as a function of right ascension (J2000).  The SDSS SN sample spans an equatorial strip, and thus the peculiar velocity correction is systematic with right ascension.  The CMB dipole correction is shown in black, while the local dipole correction, which is more relevant for low-redshift sources, is shown as shaded points, with different shades representing different redshift ranges.   \vspace{5mm} }
\label{fig:pecvel}
\end{figure}

The off-diagonal part of $C_{ij}$, given by $C_{ij}^{\rm vel}$ in Eq.~\ref{Cfull}, accounts for the
effects of correlated peculiar flows.
Expressing this quantity in an observer-centric form \citet{hui06} show that for a flat universe, 
\begin{eqnarray}
\label{cijvellarge}
&& C_{ij}^{\rm vel} =  \left[{5 \over {\,c\ln\,} 10}\right]^2
\left[1 - {a_i \over a'_i} {c \over \chiR_i} \right] 
\left[1 - {a_j \over a'_j} {c \over \chiR_j}\right] \times  \\ \nonumber 
&& \quad \quad \quad D'_i D'_j 
\int_0^\infty {dk \over 2\pi^2} \Pkzero \times \\ \nonumber 
&& \quad \quad \quad \sum_{\ell=0}^\infty 
(2\ell + 1) j'_\ell (k\chiR_i) j'_\ell (k\chiR_j) \mathcal{P}_\ell ({\bf \hat x_i} \cdot {\bf \hat x_j}),
\end{eqnarray}
where $\mathcal{P}_\ell$ is the Legendre polynomial, $j_\ell$ is the spherical Bessel function, and $j'_\ell$ is its derivative with respect to its argument.  It is useful to note that $j'_\ell(x) = j_{\ell-1} - (\ell + 1) j_\ell/x$.
This observer-centric form can be derived from Eq.~22, D7, and D10 of \citet{hui06},
by setting the survey geometry to be two delta functions localized
at the two SNe of interest. 

An alternative separation-centric form
for the same quantity is \citep{gorski88,gordon07},
\begin{eqnarray}
\label{cijvellargeB}
&& C_{ij}^{\rm vel} =  \left[{5 \over {c\,\rm ln\,} 10}\right]^2
\left[1 - {a_i \over a'_i} {c \over \chiR_i} \right] 
\left[1 - {a_j \over a'_j} {c \over \chiR_j}\right]  \times \\ \nonumber 
&& \quad \left[\, ({\bf \hat x_i} \cdot {\bf \hat r})
({\bf \hat x_j} \cdot {\bf \hat r}) \Pi (r) + 
[{\bf \hat x_i} \cdot {\bf \hat x_j} - ({\bf \hat x_i} \cdot {\bf \hat r})
({\bf \hat x_j} \cdot {\bf \hat r})] \Sigma (r) \,\right] \\ \nonumber
&& \Pi (r) \equiv D'_i D'_j \int_0^\infty {dk \over 2\pi^2} \Pkzero \left[j_0 (kr) - {2 j_1 (kr) \over kr}\right] \\ \nonumber
&& \Sigma (r) \equiv D'_i D'_j \int_0^\infty {dk \over 2\pi^2} \Pkzero {j_1 (kr) \over kr}
\end{eqnarray}
where the comoving separation between the two SNe is given by $r$ and
${\bf \hat r}$ is the unit vector pointing along the separation.  

That Eq.~(\ref{cijvellarge}) and (\ref{cijvellargeB}) are equivalent
is shown in Appendix B.
The separation-centric form is useful for fast computation,
while the observer-centric form is more directly linked to
the observed velocity angular power spectrum.
Note that these two equations are only strictly valid for a flat universe, since the derivation in Appendix B uses a plane-wave expansion that needs modification if the universe is not flat. 

We are interested in the implications of deviating from the common practise of assuming all velocities are uncorrelated.  In Section~\ref{sect:corrimpact} we therefore test the impact of including $C_{ij}^{\rm vel}$ in the covariance matrix of the uncertainties for the supernova sample used by K09.


\begin{figure}
\plotone{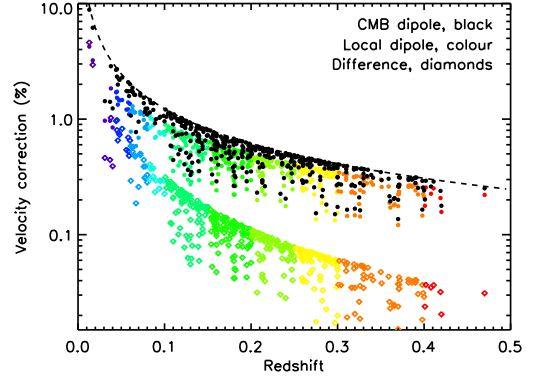}
\caption{Similar to Figure~\ref{fig:pecvel}, but with the correction expressed as a percentage and plotted against redshift.  The peculiar velocity correction taking into account the local dipole (shaded filled circles) is compared to the CMB dipole corection (black filled circles).  The difference between the two is shown as shaded diamonds.  The dashed line shows the theoretical maximum correction that would be applied to an object directly aligned with the CMB dipole: a constant peculiar velocity correction of 371\kms\ that decreases with redshift only because it is a decreasing fraction of the total redshift. The difference between the CMB and local dipole correction is only significant for the closest supernovae in the sample, with a $\sim$4\% correction in redshift when $z<0.02$ but a correction of less than 1\% for supernovae with $z\gsim0.05$.  \vspace{5mm} }
\label{fig:pecvel-percent}
\end{figure}


\begin{figure*}\bctr
\plottwo{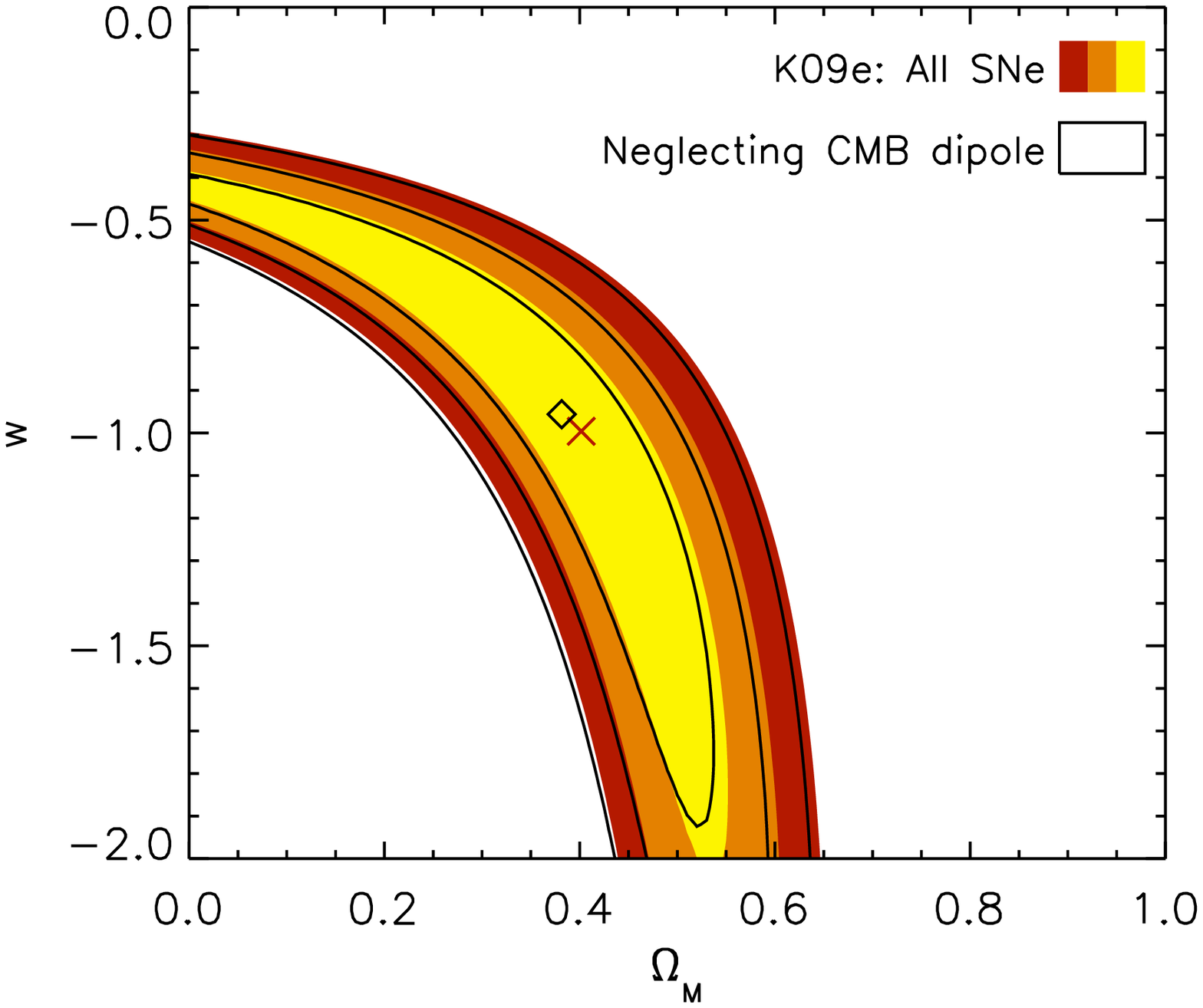}{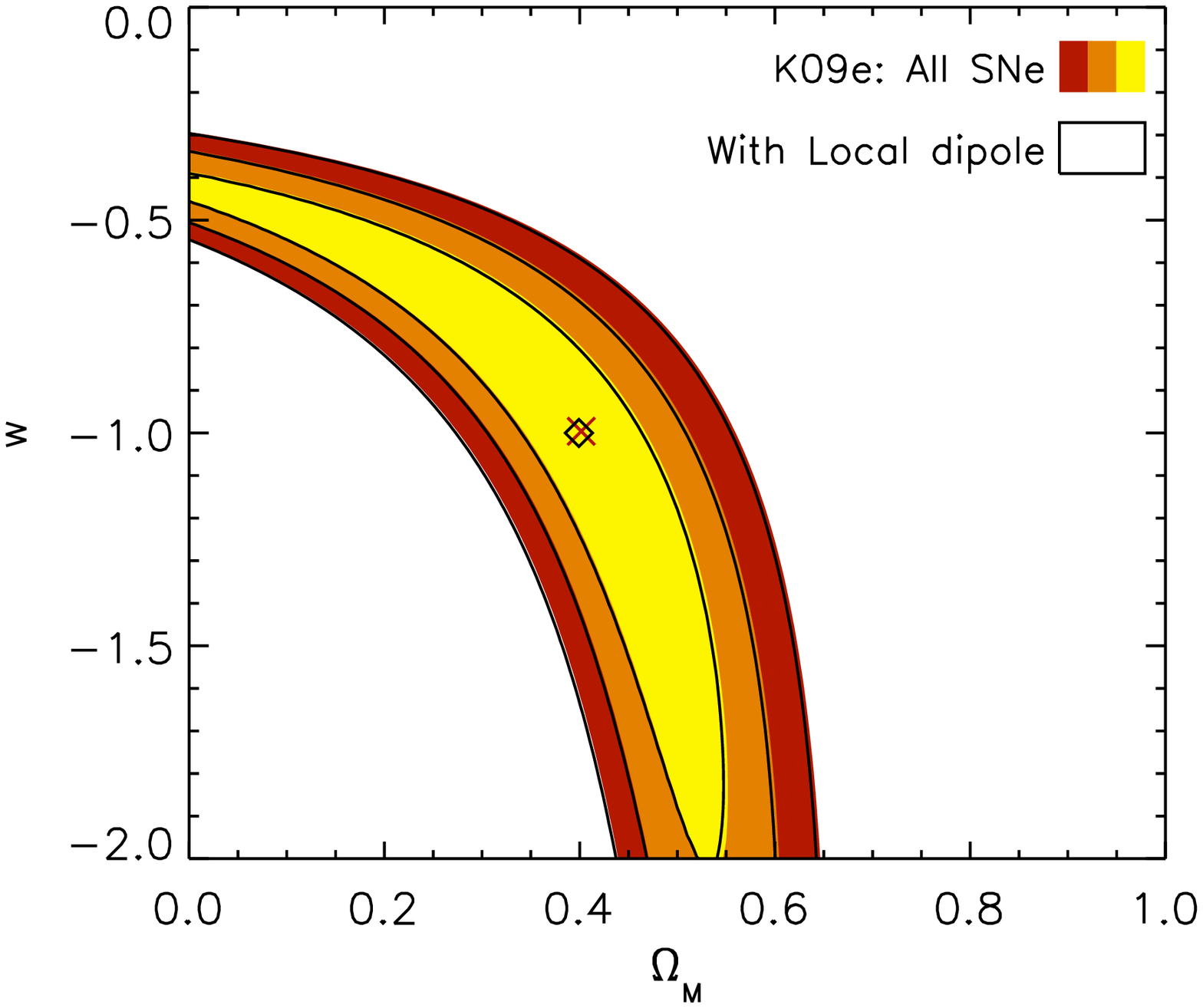}
\caption{Cosmological parameter likelihood surfaces in the flat $w$CDM model (one, two, and three sigma).  The shaded contours display the K09 data set e (full sample) and the same contours are displayed in both panels for reference.  The red cross indicates the point of maximum likelihood for the K09 data.  In addition, the left panel shows an alternative analysis of the same data in which the CMB dipole correction is not applied (black curves).  The right panel shows an alternative analysis where the both the CMB dipole and local dipole corrections are applied to the SN data.  The point of maximum likelihood for each of these alternative analyses is indicated by the black diamonds.  Only shifts perpendicular to the long axis of the contours are significant, because shifts along the long axis represent very small changes in $\chi^2$ and are well constrained by other measurements (e.g. CMB and BAO).  Correcting for the CMB dipole shifts the  contours by about 15\% of 1$\sigma$, while the shift due to the local dipole is negligible.   
}
\label{fig:contours_e}
\ectr\end{figure*}

\vspace{5mm}
\section{Impact on cosmological parameters}\label{sect:impact}

\subsection{Impact of the local dipole}\label{sect:dipoleresults}

It is inappropriate to apply the full CMB dipole correction to nearby supernovae because galaxies in our local universe share some of our locally induced peculiar motion.  For example, below redshifts of $\sim$0.02 we are strongly influenced by the Great Attractor and the Perseus-Pisces Supercluster \citep{erdogdu06}. 
\citet{bonvin06dipole} measured our dipole relative to the nearby sample of 44 supernovae used by \citep{astier06} and found it to be consistent with the CMB dipole, although with large uncertainties (about $\pm30\degree$ directional uncertainty, and 200\kms magnitude uncertainty).
\citet{haugbolle07} were able to more precisely measure the velocity flow of the local universe using the 133 low-redshift type Ia supernovae from \citet{jha07}.  At a redshift of $\sim 0.02$ (60\hinvMpc) they find a dipole amplitude of $239^{+70}_{-96}$~\kms 
in the direction $\ell\approx 281\degree\pm23\degree$, $b\approx14\degree\pm16\degree$ (measured relative to the CMB rest frame).  In Figure~\ref{fig:map} this dipole is marked by diamonds.  The magnitude of this dipole decreases with redshift. 

This result is consistent with a recent compilation by \citet{watkins09} who combined nine peculiar velocity datasets measured using five different methods of distance estimation (surface brightness fluctuations; fundamental plane; type Ia supernovae; Tully-Fisher; and brightest cluster galaxies) and concluded that the bulk flow within a Gaussian window of radius 50\hinvMpc\, is $407\pm81$\kms\, toward $\ell=287\degree \pm9\degree$, $b=8\degree \pm 6\degree$.  They note that the magnitude of this flow is larger than would be predicted by standard cosmological models based on the best cosmological parameter estimates from WMAP \citep{komatsu09}.  An even more significant deviation from the predictions of $\Lambda$CDM was found by \citet{kashlinsky08,kashlinsky09}, who measured the kinematic Sunyaev-Zel'dovich effect of the CMB in the direction of known galaxy clusters.  Their observations were at higher redshift ($z\sim0.1$ or about $300$\hinvMpc) and they found a considerably higher amplitude bulk flow (600-1000\kms) than measured by \citet{watkins09} and \citet{haugbolle07}, but the dipole direction was the same.  As well as differing from the theoretical prediction, this result does contradict the measurement by \citet{haugbolle07} of a dipole amplitude that decreases with distance.

Given the uncertainty in the redshift dependence of these local motions it is premature to apply any correction before publishing observational data.  Nevertheless we want to estimate their effect on cosmological inferences.  Since the low-redshift supernova data is the most influenced by redshift uncertainties, we estimate local motions by a local dipole according to the consistent measurements of \citet{haugbolle07} and \citet{watkins09} at around $z\sim0.015$.  We then choose to use the redshift dependence measured by  \citet{haugbolle07}, in which the dipole magnitude decreases with distance.  This relation is qualitatively what is predicted by standard $\Lambda$CDM models; however, quantitatively it does not drop as quickly as one expects from simulations.  Our choice of redshift dependence therefore lies between the theoretical predictions and the kinematic SZ results.  The precise choice is not significant because it is the low-redshift points that have the largest $dm/dz$ effect.

The magnitude of the correction due to the local dipole is shown for the SDSS supernovae as a function of right ascension in Figure~\ref{fig:pecvel} and as a function of redshift in Figure~\ref{fig:pecvel-percent}.  The black points show the correction for the CMB dipole, ignoring the local dipole.  The shaded points show the correction after the contribution from the local-dipole has been included to account for our lower velocity relative to nearby galaxies than relative to the CMB.  The lowest redshifts receive the largest local-dipole correction, while the higher-redshifts tend towards the CMB correction.


\begin{figure*}\bctr
\plottwo{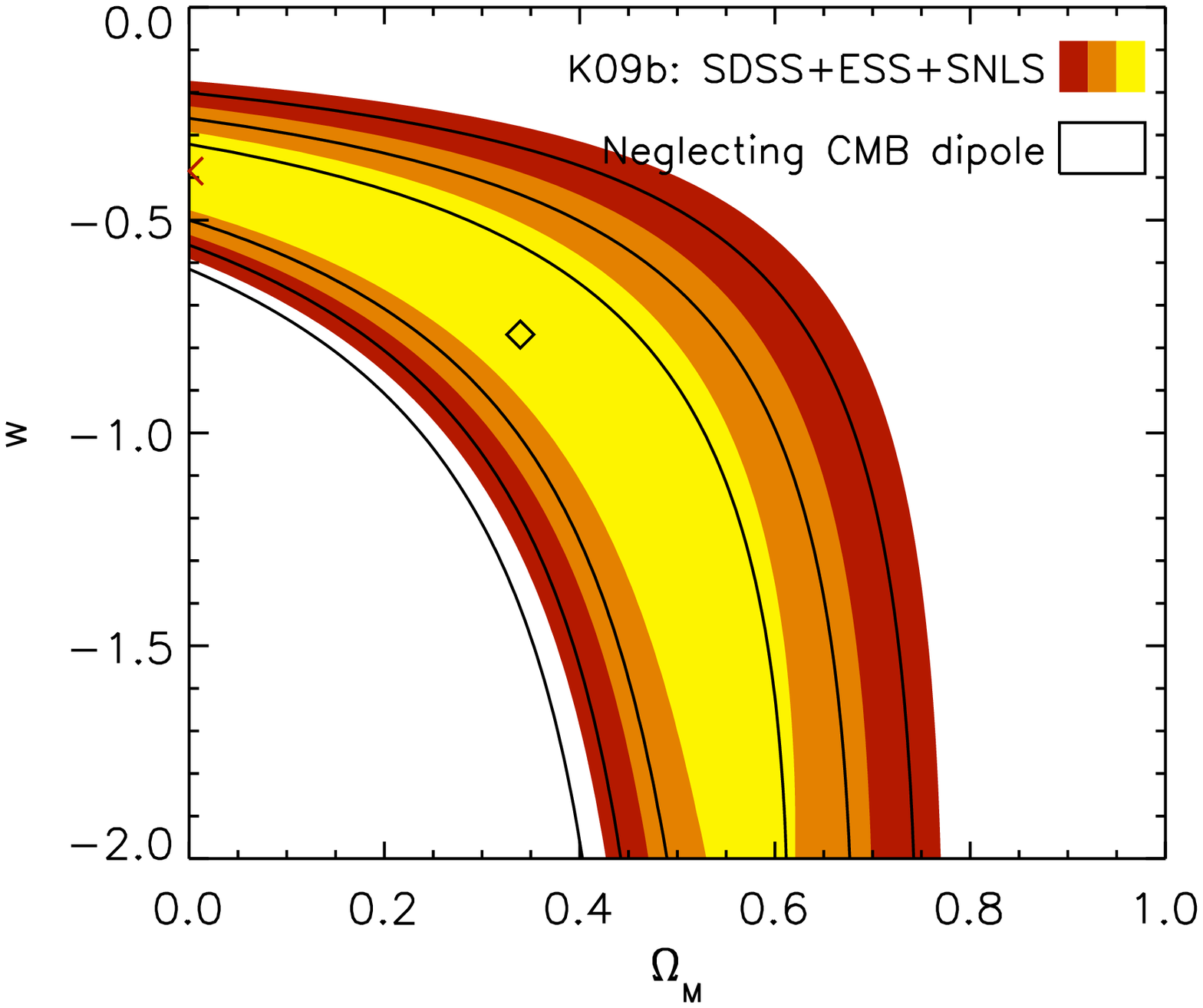}{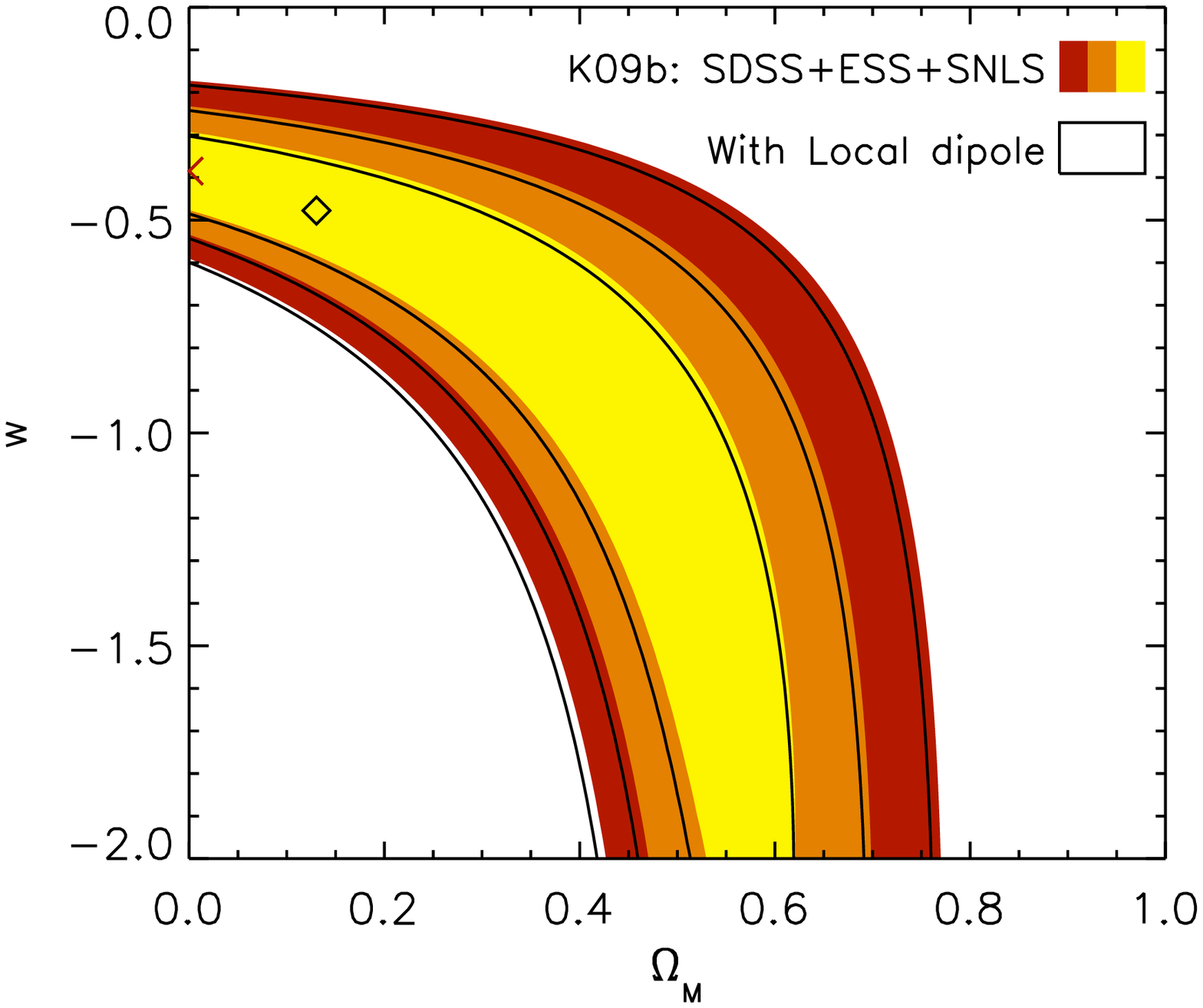}
\caption{Similar to Fig.~\ref{fig:contours_e}, but using dataset `b' from K09, which includes the SDSS, ESSENCE, and SNLS data.   This dataset is more sensitive to dipole corrections because it excludes the relatively isotropically distributed nearby sample and relies on the SDSS sample as the local anchor of the Hubble diagram.  The best fit values are shown (as red crosses for K09, black diamonds for the two variations), but these are not particularly good indicators because the best fit in this case is on the edge of the parameter space explored.  More indicative of the magnitude of the effect is the amount that the contours shift.  The CMB dipole shifts the  contours by about 0.2$\sigma$, corresponding to a $\Delta w \sim 0.04$ along the line of constant $\OM=0.3$, while the shift due to the local dipole is small ($\Delta w \lsim 0.01$).   
}
\label{fig:contours_b}
\ectr\end{figure*}

In Figure~\ref{fig:contours_e} we show the shift in cosmological parameter estimates that occurs when we ignore the CMB dipole correction and when we add the additional correction due to the local dipole.  When we simulate the effect of these dipoles we change both the observed redshift and observed luminosity distance of the supernovae, according to Eq. 6 and 11 respectively.

To make this comparison we have used the MLCS2k2 version of data set `e' from K09, which includes the new supernovae from the SDSS collaboration combined with the high-redshift ESSENCE, SNLS, and HST data along with the low-redshift sample \citep{hamuy96,riess99,jha06}.  

The choice of dataset is important, so we also show in Figure~\ref{fig:contours_b} the effect of the dipole on data set `b' from K09, which {\em excludes} the relatively isotropically distributed nearby sample and relies on the SDSS sample as the nearby anchor for the Hubble diagram.  This makes it more sensitive to the dipole correction.

It is clear from these figures that the CMB dipole correction can be important.  Neglecting it introduces a systematic shift of $\Delta w = 0.04$ for data set `b' when considering the best fit $w$ at a constant $\OM\sim0.3$.  This represents about 20\% of one standard deviation at the current accuracy levels.   This CMB correction is already routinely applied to supernova data sets and it can be done to very high precision thanks to the accurate measurements of the CMB dipole.  However, the characteristics of the local dipole are much more uncertain.  Given the assumptions we have outlined here, Figure~\ref{fig:contours_e} shows the contribution from the local dipole is currently negligible, giving a shift of $\Delta w \lsim 0.01$.  For the current sample, therefore, K09 are justified in correcting solely for the CMB dipole and ignoring local non-CMB contributions.  This may not be true in the future when more data are available, particularly data at redshifts below 0.05.  

The lower sensitivity to dipole corrections shown in  Figure~\ref{fig:contours_e} compared to Figure~\ref{fig:contours_b} demonstrates that choosing an isotropically distributed local supernova sample protects us, to a great extent, from systematic errors due to any unaccounted-for local dipole, because including all directions increases the scatter about the Hubble relation without adding bias.  In the next section we will see that isotropic samples do not save us from the effects of higher-order motions.

\subsection{Impact of correlated velocities}\label{sect:corrimpact}

We use a model linear power spectrum and growth function based on a fiducial flat-$\Lambda$CDM cosmology with $[h, \om, \ob, \se, n]=[0.701, 0.2792, 0.046, 0.817, 0.96]$ to estimate the covariance matrix for the K09 sample, $C_{ij}$, as per Eq.~\ref{cijvellarge} (see Fig.~\ref{fig:Cv}).  This encodes the likelihood that two supernovae will have correlated velocities based on their physical separation.   

Using this covariance matrix, rather than the usual uncorrelated error estimates, we re-fit our cosmological models.  We fit a flat $w$CDM model, allowing the matter density $\om$ and dark energy equation of state, $w$, to vary.  (Note that we do not redo the velocity covariance approximation for each different model, but the differences would be small.)   To the supernova fits we add the same additional observations as K09, described in detail in their Section~8.  Specifically, for Baryon Acoustic Oscillations (BAO) we use the \citep{eisenstein05} result that the derived distance parameter, $A(z=0.35)=0.469\pm0.017$, and from the CMB we use the \citep{komatsu09} result that the shift parameter $R(z=1100)=1.710\pm0.019$.  

The low-redshift cutoff is usually applied in order to remove the effect of low-redshift peculiar velocities.  Implementing a correlated velocity correction increases the error bars on all the low-$z$ supernovae relative to the high-$z$ supernovae.  This effectively down-weights the lower redshift end of the Hubble diagram and thus has a similar effect to the low-redshift cutoff. 

We first consider only the low-redshift supernovae, for which this gives the largest effect, i.e. the Low-$z$+SDSS sample `c' from K09.  As done in K09, we apply a low-$z$ cut of $z_{\rm cut}=0.02$.  In this case $w$ increases by 0.02 when the distance correlations are included.  
Using the larger sample `d' from K09 that also includes the higher redshift ESSENCE and SDSS data, we find $w$ increases by 0.014 due to the correlated velocities.  These results should be compared to the uncertainty on $w$ of $\pm0.07{\rm (stat)} \pm0.11{\rm (sys)}$ reported by K09.  Thus neglecting coherent velocities represents a potential systematic error on the best-fit value of $w$ of up to about 2\%, or about 13\% of the current estimated systematic error budget.\footnote{Adding 0.014 to the systematic error budget represents an increase of 13\% over the current 0.11 systematic uncertainty estimate.}  When future supernova surveys achieve (supernova only) statistical error bars less than about 2\%, this potential systematic error will need to be considered carefully, especially for surveys with many nearby supernovae. Indeed the Carnegie supernova project have already found that the magnitude scatter in their sample of $z<0.08$ supernovae is limited by peculiar velocities \citep[see][Fig.~19]{folatelli10}. 


\begin{figure}\begin{center}
\plotone{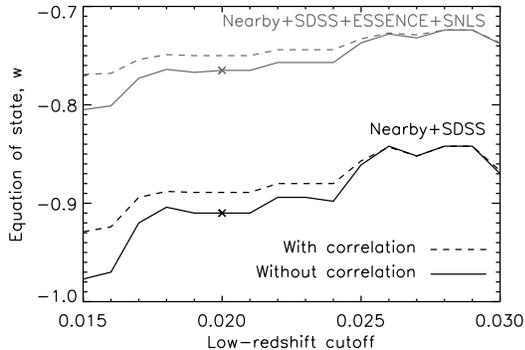}
\caption{\footnotesize Effect of low-$z$ cut on SN data, where the solid lines show the best fit equation of state obtained while neglecting correlated errors and the dashed lines show the same when correlations have been taken into account statistically (model assumes flatness with $w$ and $\om$ the only free parameters, c.f.~K09 Fig.~21).  The upper and lower sets of lines show how much the results differ when you respectively include or exclude high-redshift data.  Upper (grey) curves are for the K09 data set `d', with Nearby, SDSS, ESSENCE, and SNLS supernova samples.  Lower (black) curves are for the K09 data set `c', with only the Nearby and SDSS data included.  K09 use a low-redshift cutoff of 0.02 (crosses) with systematic uncertainties (calculated for the flat-$\Lambda$CDM model with the MLCS light-curve fitter) of $\pm0.11$ for set `d' and $+0.10-0.33$ for set `c' (the greater systematic uncertainty in the lower direction for the set `c' arises primarily due to uncertainties in the rest-frame $U$ band).  So the offset seen here between these two data set combinations is within the uncertainties.  The effect of correlations is currently smaller than the other systematic uncertainties considered in K09, but will be important for attempts to measure $w$ to better than 3\%.  Raising the low-redshift cutoff to 0.025 is sufficient to remove the expected effect of correlated supernova motions.}
\label{fig:corr-effect-on-w}
\end{center}\end{figure}

Figure~\ref{fig:corr-effect-on-w} shows the effect of implementing a range of different low-redshift cutoffs on the supernova data, both for the original K09 MLCS2k2 data (solid lines) and for our version of that data with uncertainties corrected for correlated motion (dashed lines).\footnote{Note that K09 also uses the SALT II light-curve fitter and the results differ.  We do not debate the merits of different light-curve fitters here, our qualitative results are relevant whichever light-curve fitter is used.}  We plot the best fit value of $w$ derived for a flat model with $w$ and $\om$ as free parameters.  

The correlation-corrected result (dashed line) can be matched by implementing a larger low-$z$ cut on the uncorrected data (e.g.\ the correlation-corrected result with $z_{\rm cut}\approx0.015$ matches the uncorrected data with $z_{\rm cut}\approx0.017$).
 It is also evident that the effect of the low-$z$ cut on the data with correlated errors is smaller than on the data set with uncorrelated errors.  Both of these features are as expected, because some of the low-$z$ cut was already effectively implemented by the down-weighting due to correlations.  We also note that the SDSS supernovae are much less prone to correlations than the Nearby sample, simply due to their greater physical distances between the supernovae.  SDSS supernova therefore provide a larger improvement to the low-redshift anchor of the magnitude-redshift diagram than one might na\"{i}vely expect.
  
This analysis demonstrates that if the velocity covariance matrix had been used in K09 then the need for a low-$z$ cut would have been diminished.   
Using the covariance matrix for the peculiar velocity uncertainties should be an optimal statistical treatment of the supernova data.  It automatically includes the effect of monopole uncertainties and dipole uncertainties as well as the higher-order correlated motions.  Although slightly more complicated to implement than a simple low-$z$ cut, the results are more robust.  

Computing the full correlation matrix does have some disadvantages, primarily because it is model dependent.  Our calculation of correlations has been made in a fiducial $\Lambda$CDM model, 
so it is not strictly self-consistent to use these correlations to test other models.   However, this is mitigated by the fact that the majority of the covariance signal comes from low redshifts, and most viable models for the universe have to agree fairly closely on the evolution and growth of structure in the local universe in order to match observations.  To check the differences are negligible the correlations can be self-consistently re-derived for each cosmological being fitted. 

As future surveys with many more supernovae attempt to obtain percent-level accuracy on the value of the equation-of-state parameter the effect of correlations will become ever more important.   At the very least, neglecting correlations under-estimates the {\em uncertainty} on our cosmological inferences, and in the worst-case scenario can bias the values of cosmological parameters we derive.


\begin{figure*}\bctr
\plottwo{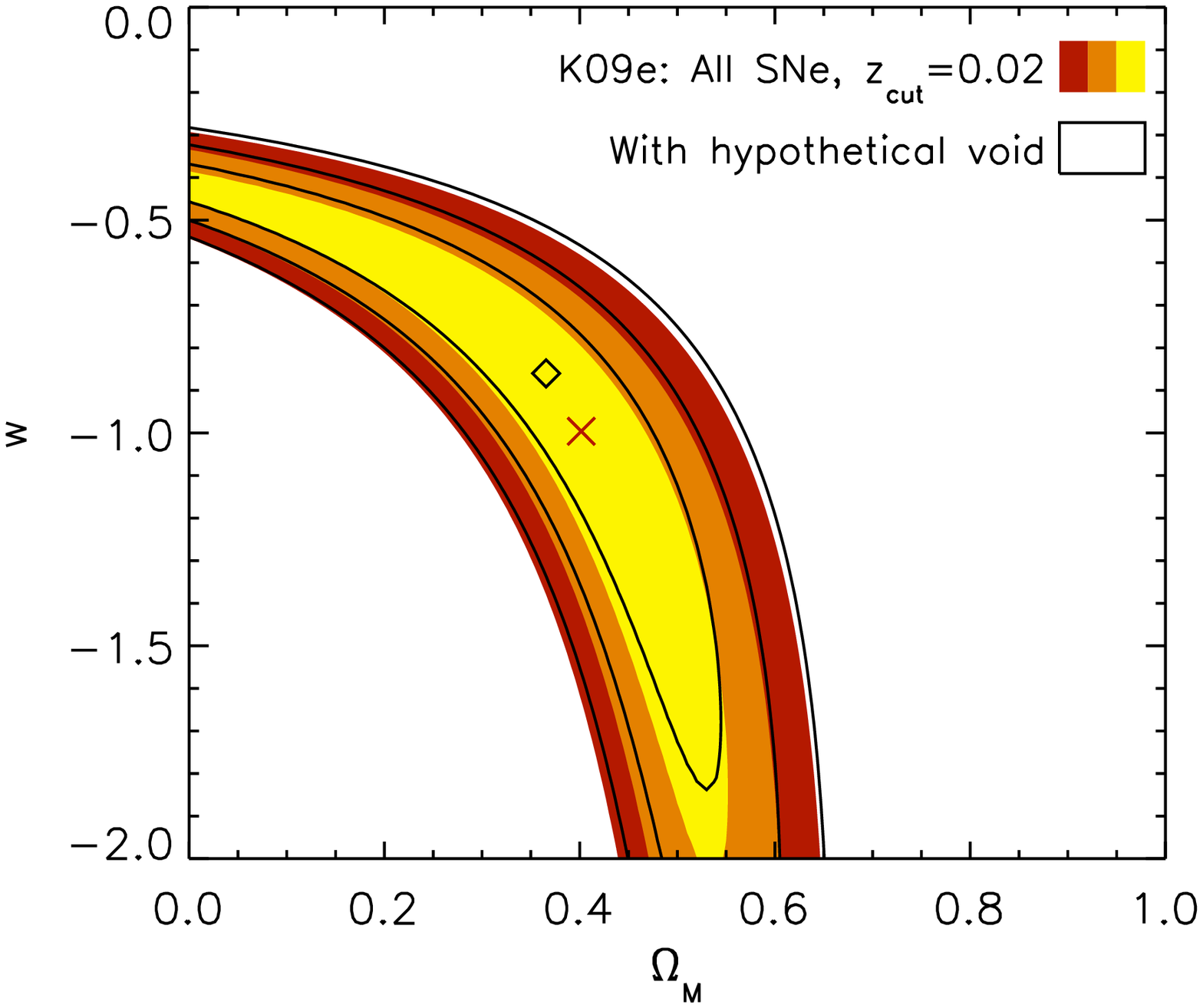}{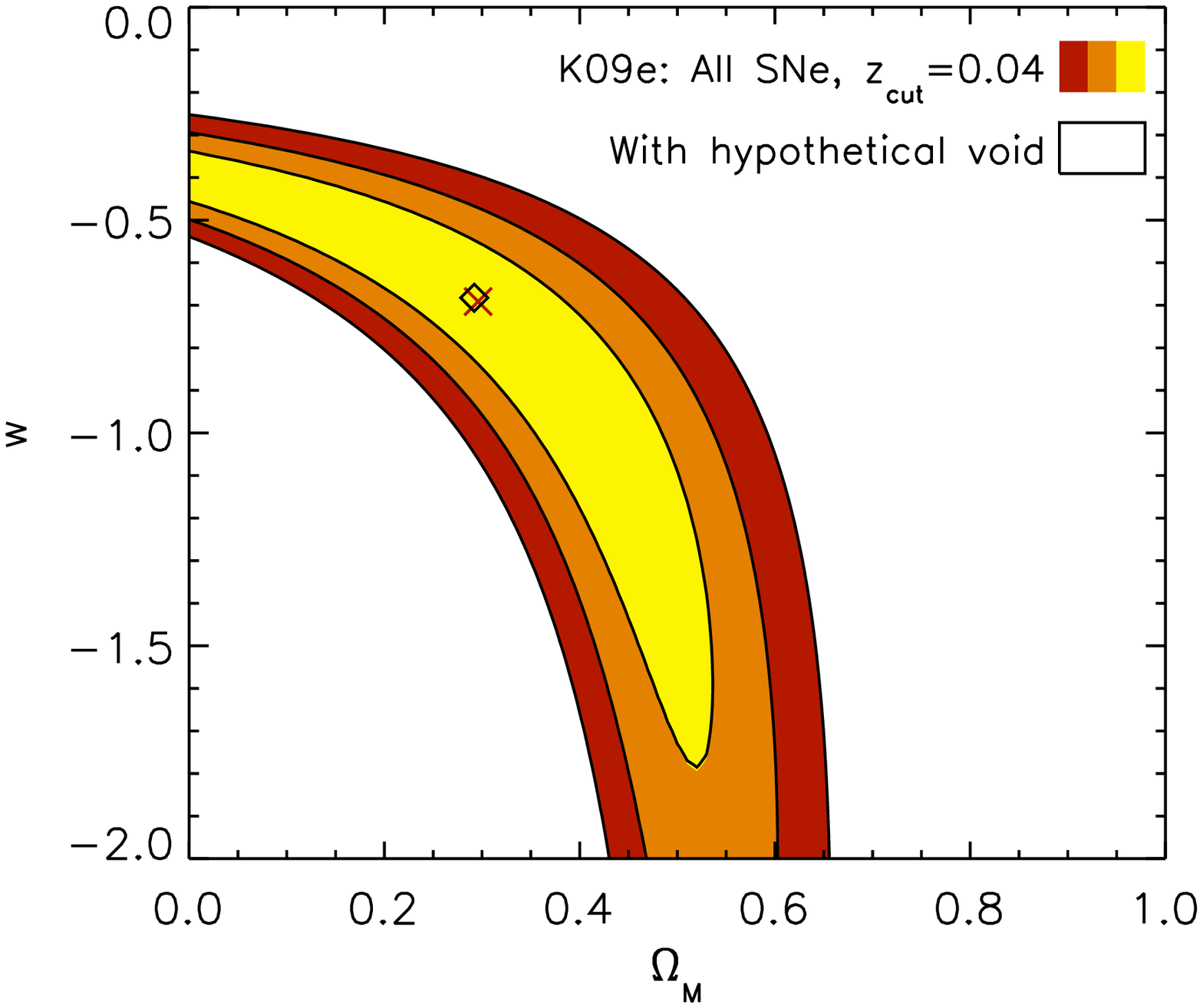}
\caption{This figure demonstrates the effect of correcting the SN data for a hypothetical `Hubble Bubble', in this case a Gaussian underdensity of 30\% on a scale of 70\hinvMpc.  In the left panel the shaded contours display the K09 data set e (with the maximum likelihood indicated by the red cross) and black outline indicates the results for the same data corrected for a local void (with the maximum likelihood indicated by a black diamond).  In the left panel the low-redshift cutoff was the standard $z_{\rm cut}=0.02$.  
The right panel demonstrates the effect of increasing the low-redshift cutoff from $z_{\rm cut}=0.02$ to $z_{\rm cut}=0.04$.  The analysis is identical to the left panel except that in both the homogeneous case (shaded contours) and the putative `Hubble bubble' case (black outline), the low-redshift cutoff was $z_{\rm cut}=0.04$.  Although increasing the low-redshift cut reduces the susceptibility of the data to local density fluctuations, dropping the low-redshift data changes the best fit cosmology by more than the local void would because of the weakened constraints.  See also how the best fit $w$ changes with changing low-$z$ cut in Fig.~\ref{fig:corr-effect-on-w}.  }
\label{fig:contours_void}
\ectr\end{figure*}

\vspace{5mm}
\section{Impact of Local under-density}\label{sect:monopole}
Peculiar velocities are not the only source of systematic redshift effects on nearby galaxies.  
The potential presence of a systematic redshift due to our possible position at the center of a local underdensity is of particular interest to supernova cosmology because the discovery of the acceleration of the universe is founded on the observation that high-redshift type Ia supernovae appear to be more distant than expected in a decelerating universe \citep{riess98,perlmutter99}.  This observation has incited intense scrutiny of the potential for a local underdensity, known as a ``Hubble bubble'', that may be influencing our results \citep{zehavi98,jha06,conley07}, but the massive size needed to explain away the acceleration \citep{alnes06a,enqvist07,enqvist08,garcia-bellido08,garcia-bellido08-kz,garcia-bellido09,zibin08,clifton08,alexander09,marra10} and the need to be almost at the center of such an underdensity \citep{alnes06b,caldwell08,blomqvist10} makes such a model very contrived. 

Observations of local large scale structure indicate that we do sit in a local underdensity \citep[e.g.][]{geller97,gottloeber10}, albeit much smaller than that needed to explain a cosmological constant (on the order of 100 Mpc, as opposed to 1 Gpc).  In Appendix C we use numerical simulations to show that underdensities of this size are the size and depth of typical density fluctuations in a $\Lambda$CDM model.  It is therefore important to investigate the impact that these realistic density fluctuations would have on our derivation of cosmological parameters.  

\citet{sinclair10} used the Lema\^itre-Tolman-Bondi (LTB) model to simulate a local underdensity, as outlined in \citet{garcia-bellido08} and fit for a homogeneous cosmological model.  They found that {\em if} we live in a 30\% underdensity of scale 70\hinvMpc, {\em then} assuming the universe is homogeneous could dupe us into believing that $w$ is far more phantom-like ($<-1$), and the density of dark energy far more significant, than they really are (up to a 10\% error in $w$ in the worst-case scenario, even with a low-redshift cut that excises the local underdensity).  

Given that these discrepancies remain significant it is worth investigating the potential impact on recent data sets and whether a higher low-$z$ cutoff might be worthwhile. 
To that end we here apply the \citet{sinclair10} result to the SDSS data explicitly.  At each observed redshift we alter the SDSS data for the effects of a hypothetical void of the type described above ($r_0=70$\hinvMpc, with $\delta = -0.3$).  The difference is shown in the left panel of Fig.~\ref{fig:contours_void}, in which the standard low-$z$ cut of $z_{\rm cut}=0.02$ has been used.  

Although in Fig.~\ref{fig:contours_void} there appears to be a large shift in the best fit parameters, from $(\om,w_0)=(0.40,-1.00)$ to $(0.37,-0.86)$, the final cosmological parameters are not susceptible to this full discrepancy because most of the variation is along the long-axis of the contours, which is well constrained by other observations such as the CMB and BAO.  The direction that the supernovae constrain most tightly is only changed a very small amount by the presence of a local void.  

It could nevertheless be argued that a higher low-$z$ cutoff may be necessary to avoid the effects of local inhomogeneities.  However, by excluding low redshift data one sacrifices constraining power.  In the right panel of Fig.~\ref{fig:contours_void} we show the result of increasing the low-$z$ cutoff to $z_{\rm cut}=0.04$.  Although the effect of the void is now negligible (shaded and black contours overlap) the contours have shifted further due to the loss of constraining power than they did due to the hypothetical void.  We therefore conclude that increasing the redshift cutoff is counter-productive with the size of the low-$z$ SN sample used in K09.  

This analysis strengthens the argument for using the covariance matrix approach to down-weighting low-$z$ supernovae since it inherently takes into account potential monopole velocities.

\vspace{5mm}
\section{Conclusions}\label{sect:conclusion}

From this study we can conclude that the cosmological results derived by the SDSS SN survey are robust to peculiar velocity systematics.  The local dipole represents a negligible addition to the CMB dipole correction that has already been implemented. Future surveys with many nearby supernovae may need to take it into account, but we note that an isotropically distributed local supernova sample would shield us, to a great extent, from systematic errors due to the local dipole. 

Neglecting correlated peculiar velocities can cause an error in the best-fit value of $w$, which in the current sample underestimates $w$ by about 2\%.  It also causes us to overestimate the precision of our measurement.  As future surveys aim for percent-level accuracy on the value of the equation-of-state parameter, the importance of correlations between the peculiar velocities of supernovae will increase.  Here we treated them in a statistical sense, but it may be possible in the future to correct the supernova velocities for measured local flows.  A future method of testing cosmological parameters will be to use the peculiar velocities as signal rather than noise, and generate a peculiar velocity power spectrum to compare against cosmological models.  
 In the meantime, we find that accounting for peculiar velocities by using a covariance matrix for the correlated errors is a more robust way to down-weight low-redshift supernovae than applying a sharp low-redshift cut.  Doing so does not degrade the uncertainties in $w$, despite the down-weighting of the signal, because one can include more low-redshift supernovae in the overall fit.

Finally, we used $n$-body simulations to gauge the likely distribution of local under- and over-densities and found that a density fluctuation of 30\% from the mean cosmological density, out to a range of 70\hinvMpc, is reasonable given the expectations of concordance $\Lambda$CDM.  A density fluctuation of this size can have a significant impact on the cosmological parameters we derive.  The worst of those systematic errors can be avoided by down-weighting nearby supernovae and we demonstrated that the currently used $z_{\rm cut}=0.02$ is well justified.  However, we advocate including the velocity covariance directly in one's likelihood analysis as a more systematic way to down-weight the low-$z$ SNe.

In summary, peculiar motions and gravitational effects due to inhomogeneities have systematic effects on the measurement of cosmological parameters using luminosity distance indicators such as type Ia supernovae.  These effects will become significant for the next generation of surveys and here we have suggested a covariance-matrix approach to correcting for them statistically, based on the expected correlation between the motion of the sources.  This improves on the usual low-redshift cut approach, and we have provided code that generates the covariance-matrix for any sample of supernovae, to make this technique easy to implement.

\acknowledgments
{\small 
TMD thanks Alexandra Abate for useful comments and also the Centro de Ciencias de Bensque ÒPedro PascualÓ for providing us with such a superb facility during some of the work on leading to this paper. 

Computer time was provided by the Danish Center for Scientific Computing (DCSC).  Some of the results in this paper have been derived using the HEALPix\footnote{\url{http://healpix.jpl.nasa.gov}} \citep{gorski05} package.  

Funding for the SDSS and SDSS-II has been provided by the Alfred P. Sloan Foundation, the Participating Institutions, the National Science Foundation, the U.S. Department of Energy, the National Aeronautics and Space Administration, the Japanese Monbukagakusho, the Max Planck Society, and the Higher Education Funding Council for England. The SDSS Web Site \hbox{is {\tt http://www.sdss.org/}.}

The SDSS is managed by the Astrophysical Research Consortium for the Participating Institutions. The Participating Institutions are the American Museum of Natural History, Astrophysical Institute Potsdam, University of Basel, University of Cambridge, Case Western Reserve University, University of Chicago, Drexel University, Fermilab, the Institute for Advanced Study, the Japan Participation Group, Johns Hopkins University, the Joint Institute for Nuclear Astrophysics, the Kavli Institute for Particle Astrophysics and Cosmology, the Korean Scientist Group, the Chinese Academy of Sciences (LAMOST), Los Alamos National Laboratory, the Max-Planck-Institute for Astronomy (MPIA), the Max-Planck-Institute for Astrophysics (MPA), New Mexico State University, Ohio State University, University of Pittsburgh, University of Portsmouth, Princeton University, the United States Naval Observatory, and the University of Washington.

This work is based in part on observations made at the 
following telescopes.
The Hobby-Eberly Telescope (HET) is a joint project of the 
University of Texas at Austin,
the Pennsylvania State University,  Stanford University,
Ludwig-Maximillians-Universit\"at M\"unchen, and 
Georg-August-Universit\"at G\"ottingen.  
The HET is named in honor of its principal benefactors,
William P. Hobby and Robert E. Eberly.  The Marcario Low-Resolution
Spectrograph is named for Mike Marcario of High Lonesome Optics, who
fabricated several optical elements 
for the instrument but died before its completion;
it is a joint project of the Hobby-Eberly Telescope partnership and the
Instituto de Astronom\'{\i}a de la Universidad Nacional Aut\'onoma de M\'exico.
The Apache Point Observatory 3.5 m telescope is owned and operated by 
the Astrophysical Research Consortium. We thank the observatory 
director, Suzanne Hawley, and site manager, Bruce Gillespie, for 
their support of this project. The Subaru Telescope is operated by the 
National Astronomical Observatory of Japan. The William Herschel 
Telescope is operated by the Isaac Newton Group on the island of 
La Palma in the Spanish Observatorio del Roque 
de los Muchachos of the Instituto de Astrofisica de 
Canarias. The W.M. Keck Observatory is operated as a scientific partnership 
among the California Institute of Technology, the University of 
California, and the National Aeronautics and Space Administration. 
The Observatory was made possible by the generous financial support 
of the W. M. Keck Foundation. 
}



{\small 
{\it Facilities:} \facility{Nickel}, \facility{HST (STIS)}, \facility{CXO (ASIS)}.
}


\appendix

\section{Appendix A: Treatment of random peculiar velocity contributions}\label{app:dmdz}\label{sect:random}

The motion of distant supernovae and their host galaxies imprints a peculiar velocity error that is primarily random (as opposed to our own motion, on which Sect.~\ref{sect:dipole} concentrates).  That peculiar velocity dispersion $\sigma_v^{\rm pec}\sim300$\kms\  gives a redshift error of $\sigmazp= \sigma_v^{\rm pec}/c$ (or the special relativistic formula if the peculiar velocity was higher).  The measured redshift, $z$, is a combination of the recession and peculiar velocity contributions according to $(1+z)=(1+\zr)(1+\zp)$, where $\zr$ and $\zp$ are the recession and peculiar velocity contributions to the redshift, respectively.  Differentiating this expression to calculate the error contribution from peculiar velocities gives 
\beq \sigmaz = (1+\zr)\sigmazp + (1+\zp)\sigmazr .\eeq
We can take the error in recession velocity to be zero, $\sigmazr=0$, so the uncertainty we need to add to our redshifts to account for peculiar velocities is $\sigmaz = (1+\zr)\sigmazp$.

Previous analyses, including for example \citet{davis07}, used $\sigmaz=\sigmazp$ and so slightly underestimated the contribution from peculiar velocities at high redshifts.  Formally, the uncertainty at $z=1$ should have been double what was used, but since the proportional contribution from peculiar velocities still decreases with redshift, this only corresponds to an error of 0.26\%, as opposed to 0.13\%, and the difference is negligible for cosmology. 

We convert $\sigmaz$ into an approximate magnitude uncertainty, $\sigma_{\rm m}^{\rm pec}$, using the magnitude-redshift relation, and combine it in quadrature with the uncertainty in the measured magnitude, $\sigma_{\rm m}^{\rm meas}$, and the intrinsic magnitude dispersion, $\sigma_{\rm m}^{\rm int}$, of the supernovae.  

\begin{SCfigure}[0.45][t]
\includegraphics[width=100mm]{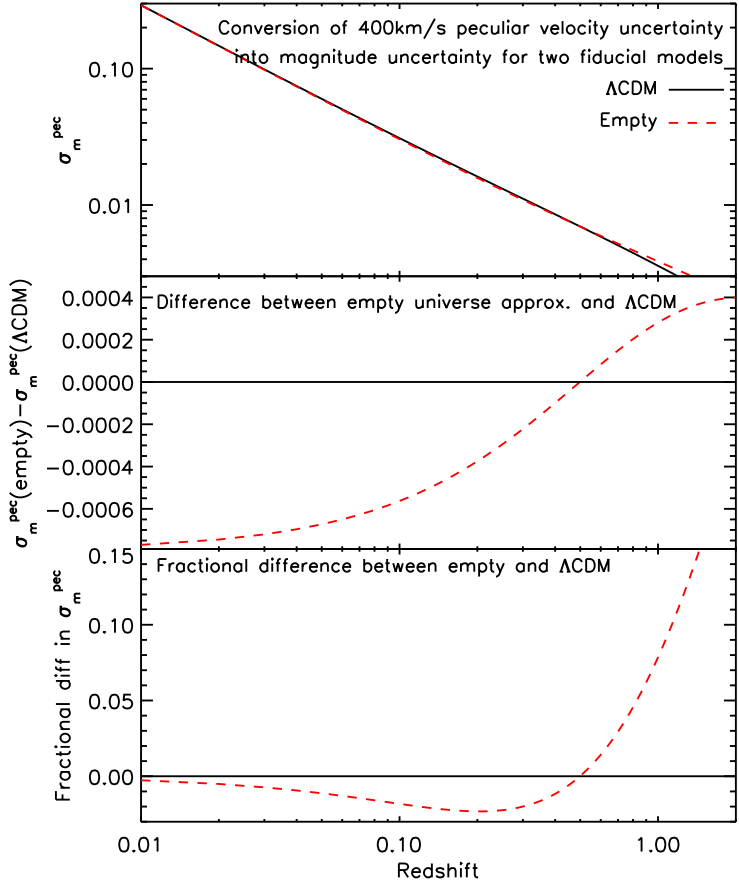}
\caption{\small Examples of the conversion from redshift uncertainty to magnitude uncertainty.  A peculiar velocity uncertainty of $\sigma_v^{\rm pec}=$300\kms\ corresponds to $\sigma_z^{\rm pec}$=0.001.  This value converts to a large magnitude uncertainty at low redshift, where the slope of the magnitude-redshift diagram is steep, but a smaller magnitude uncertainty at high redshift.  Different fiducial models give slightly different conversions between redshift and magnitude uncertainties, but the difference is negligible for cosmological inferences.  Here the empty model conversion (Eq.~\ref{eq:empty}) is compared to the $\Lambda$CDM model conversion (Eq.~\ref{eq:dmu} with $\om=0.3$, $\oll=0.7$).  In absolute terms (middle panel) the difference is largest at low redshift, but in relative terms (lower panel) the difference is largest at high redshifts.  The lower panel shows 
\begin{displaymath} 
\frac{\left[\sigma_m^{\rm pec}({\rm empty})-\sigma_m^{\rm pec}(\Lambda{\rm CDM})\right] }{ \sigma_m^{\rm pec}(\Lambda{\rm CDM})}.
\end{displaymath}  \vspace{1cm}}
\label{fig:z-mag-conversion}
\end{SCfigure}

The distance modulus has the form
\beq \mu = 5 \log_{10}(\dLbar) = \frac{5}{\ln(10)}\ln\left[\chiR(1+\zr)\right],\eeq
where $\chiR=R_0\chi=c\int_0^\zr H(z)^{-1}dz$ is the comoving distance.  
Therefore an error in the redshift corresponds to a magnitude error of
\beq \sigmamu = \sigmaz \frac{5}{\ln(10)}\left[\frac{1}{1+\zr} + \frac{c}{\chiR H(\zr)}\right]. \label{eq:dmu}\eeq 
Although a fiducial cosmology is used for this calculation -- often taken to be $\Lambda$CDM with $\OM\sim 0.3$ and $\oll\sim 0.7$ -- differences from the derived cosmology are small and have negligible impact on cosmology fits.  Note that K09 use the empty universe as their fiducial cosmology for error calculations, and approximate the empty universe case, in which $H(\zr)=H_0(1+\zr)$ and $\chiR=c\ln(1+\zr)/H_0$, by 
\beq \sigmamu \sim \sigmaz \frac{5}{\ln(10)}\left[\frac{1+\zr}{\zr(1+\zr/2)}\right]. \label{eq:empty} \eeq 
The difference between this approximation and Eq.~\ref{eq:dmu} is shown in Fig.~\ref{fig:z-mag-conversion}.

In the non-flat case $\chiR$ should be replaced with $R_0S_k(\chi)$ 
in the equation for $\mu$, or $R_0T_k(\chi)$ 
in the equation for $\sigma_\mu$, where $S_k=\sin$ or $\sinh$ in the closed and open cases, respectively, and $T_k=\tan$ or $\tanh$.


\section{Appendix B: The equivalence of the observer-centric and separation-centric expressions for the magnitude covariance matrix}
Here we derive the expressions for $C_{12}^{\rm vel}$ 
in Eq.~\ref{cijvellarge} and \ref{cijvellargeB}.  These derivations are valid in the flat universe case. 
It comes down to evaluating the two point velocity correlation 
$\xi^{\rm vel}_{12} \equiv \langle ({\bf v_1} \cdot {\bf \hat x_1}) ({\bf v_2} \cdot {\bf \hat x_2}) \rangle $, where $1$ and $2$ label the two SNe in question, since
\begin{eqnarray}
C_{12}^{\rm vel} = \left[{5 \over {c\,\rm ln\,} 10}\right]^2
\left[1 - {a_1 \over a'_1} {c \over \chiR_1} \right] 
\left[1 - {a_2 \over a'_2} {c \over \chiR_2}\right] {\xi^{\rm vel}_{12} } .\\ \nonumber
\end{eqnarray}
This is actually an old subject \citep[see e.g.][]{gorski88}. One reason we
go over the derivation here is that errors have crept into some
recent literature, as pointed out by \citet{gordon07}.
It is also useful to see how two completely different looking expressions,
i.e. Eq.~\ref{cijvellarge} and \ref{cijvellargeB}, are actually equivalent.
Errors have occurred in some recent versions of Eq.~\ref{cijvellarge}
\citep[e.g.][]{hernandez06,cooray06}.

Using linear theory, it can be shown that
\beq
\label{lineartheory}
 \xi^{\rm vel}_{12} = D'_1 D'_2 
\int {d^3 k \over (2\pi)^3} k^{-2} \Pkzero \,  
 ({\bf \hat k} \cdot {\bf \hat x_1})
({\bf \hat k} \cdot {\bf \hat x_2}) e^{-i{\bf k} \cdot ({\bf x_1} - {\bf x_2})}
\eeq
where ${\bf x_1}$ and ${\bf x_2}$ are the comoving positions of the two SNe in question,
${\bf \hat x_1}$ and ${\bf \hat x_2}$ are the unit vectors pointing in these directions,
$\Pkzero$ 
is the mass power spectrum today, and
$D'_1$ and $D'_2$ are the derivatives of the growth factor with respect to conformal time
at the two redshifts of interest.

An observer-centric approach is to use
\begin{eqnarray}
{\bf \hat k} \cdot {\bf \hat x_2} e^{i {\bf k} \cdot {x_2}} = 
4\pi \sum_{\ell, m} i^{\ell - 1} j'_\ell (k \chiR_2) Y_{\ell m}^* ({\bf \hat k}) Y_{\ell m} ({\bf \hat x_2})
\end{eqnarray}
where $j_\ell$ is the spherical Bessel function, $j'_\ell$ is its derivative (with respect 
to its argument, not conformal time), and $Y_{\ell m}$'s are the spherical harmonics.
Performing the integral over ${\bf \hat k}$ in Eq.~\ref{lineartheory}, and using
$\int d\Omega_k Y^*_{\ell m} ({\bf \hat k}) Y_{\ell' m'} ({\bf \hat k}) = \delta_{\ell \ell'} \delta_{m m'}$
and 
$\mathcal{P}_\ell ({\bf \hat x_1} \cdot {\bf \hat x_2}) = 4\pi/(2\ell+1) \sum_m Y^*_{\ell m} ({\bf \hat x_1})
Y_{\ell m} ({\bf \hat x_2})$,
it is straightforward to show that
\begin{eqnarray}
\xi^{\rm vel}_{12} = D'_1 D'_2 
\int {dk \over 2\pi^2} \Pkzero\, 
 \sum_\ell (2\ell + 1) j'_\ell (k\chiR_1) j'_\ell(k\chiR_2)
\mathcal{P}_\ell ({\bf \hat x_1} \cdot {\bf \hat x_2})
\end{eqnarray}
from which Eq.~\ref{cijvellarge} can be obtained \citep[see][for details]{hui06}.
The above expression is observer-centric in the sense that one can easily read off
from it the angular velocity power spectrum as seen by the observer,
\begin{eqnarray}
\label{velpower}
\mathcal{C}_\ell^{\rm vel} = D'_1 D'_2 \int {2 dk \over \pi} \Pkzero\,  j'_\ell (k\chiR_1) j'_\ell (k\chiR_2).
\end{eqnarray}
Here, $1$ and $2$ can refer to the same redshift, or two different redshifts.

A different approach to reducing Eq.~\ref{lineartheory} is
to first note that by symmetry arguments \citep{gorski88},
\begin{eqnarray}
\langle v_i ({\bf x_1}) v_j ({\bf x_2}) \rangle
= [\Pi (r) - \Sigma (r)] \hat r_i \hat r_j + \Sigma (r) \delta_{ij},
\end{eqnarray}
where $i$ and $j$ here, unlike in the rest of the paper, label the spatial directions
rather than the SNe, $r$ is the comving separation between points 1 and 2, and
${\bf \hat r}$ is the associated unit vector. Suppose ${\bf \hat r}$ points in
the $z$ direction, then the above matrix is diagonal, with diagonal entries
$\Sigma, \Sigma, \Pi$ i.e. $\Sigma$ is the perpendicular velocity correlation
and $\Pi$ is the parallel velocity correlation. 
Here, parallel and perpendicular are defined by the separation vector between
the two SNe (hence a separation-centric approach).
From this matrix, one can deduce that
\begin{eqnarray}
\xi_{12}^{\rm vel} = ({\bf \hat x_1} \cdot {\bf \hat r}) ({\bf \hat x_2} \cdot {\bf \hat r}) \Pi (r)\, 
 + 
[{\bf \hat x_1} \cdot {\bf \hat x_2} - ({\bf \hat x_1} \cdot {\bf \hat r})
({\bf \hat x_2} \cdot {\bf \hat r})] \Sigma (r)
\end{eqnarray}
where $[{\bf \hat x_1} \cdot {\bf \hat x_2} - ({\bf \hat x_1} \cdot {\bf \hat r})
({\bf \hat x_2} \cdot {\bf \hat r})]$ can be written as 
${\,\rm sin}\theta_1 {\,\rm sin}\theta_2$ if ${\bf \hat x_1} \cdot {\bf \hat r} =
{\,\rm cos}\theta_1$ and ${\bf \hat x_2} \cdot {\bf \hat r} = {\,\rm cos}\theta_2$.
Comparing this expression with Eq.~\ref{lineartheory}, one can see that,
\begin{eqnarray}
\Pi(r) = D'_1 D'_2 \int {d^3 k \over (2\pi)^3} k^{-2} \Pkzero 
({\bf \hat k} \cdot {\bf \hat r})^2 e^{i{\bf k} \cdot {\bf r}}.
\end{eqnarray}
Using 
\begin{eqnarray}
e^{i {\bf k} \cdot {\bf r}} = 
\sum_{\ell} (2\ell + 1) i^{\ell} j_\ell (k r) \mathcal{P}_\ell ({\bf \hat k} \cdot {\bf \hat r})
\end{eqnarray}
and integrating over ${\bf \hat k}$ (choosing ${\bf \hat r}$ to lie in the z direction
for instance), one can see that only $\ell = 2$ and $\ell = 0$ survives.
Finally, using $j_2 = 3 j_1/x - j_0$, one obtains,
\beq
\Pi(r) = D'_1 D'_2 \int {dk \over 2\pi^2} \Pkzero \,
\left[ j_0 (kr) - {2 j_1 (kr) \over kr} \right].
\eeq
The perpendicular counterpart can be similarly obtained from,
\begin{eqnarray}
\Sigma(r) = D'_1 D'_2 \int {d^3 k \over (2\pi)^3} {\Pkzero \over k^2}
({\bf \hat k} \cdot {\bf \hat x})^2 e^{i{\bf k} \cdot {\bf r}},
\end{eqnarray}
with ${\bf \hat x}$ pointing in the x direction while ${\bf \hat r}$ points
in the z direction. A few manipulations yield,
\begin{eqnarray}
\Sigma (r) = D'_1 D'_2 \int {dk \over 2\pi^2} \Pkzero\, {j_1 (kr) \over kr},
\end{eqnarray}
reproducing the results of \citet{gorski88} and giving our Eq.~\ref{cijvellargeB}.


\begin{figure*}
\includegraphics[width=84mm]{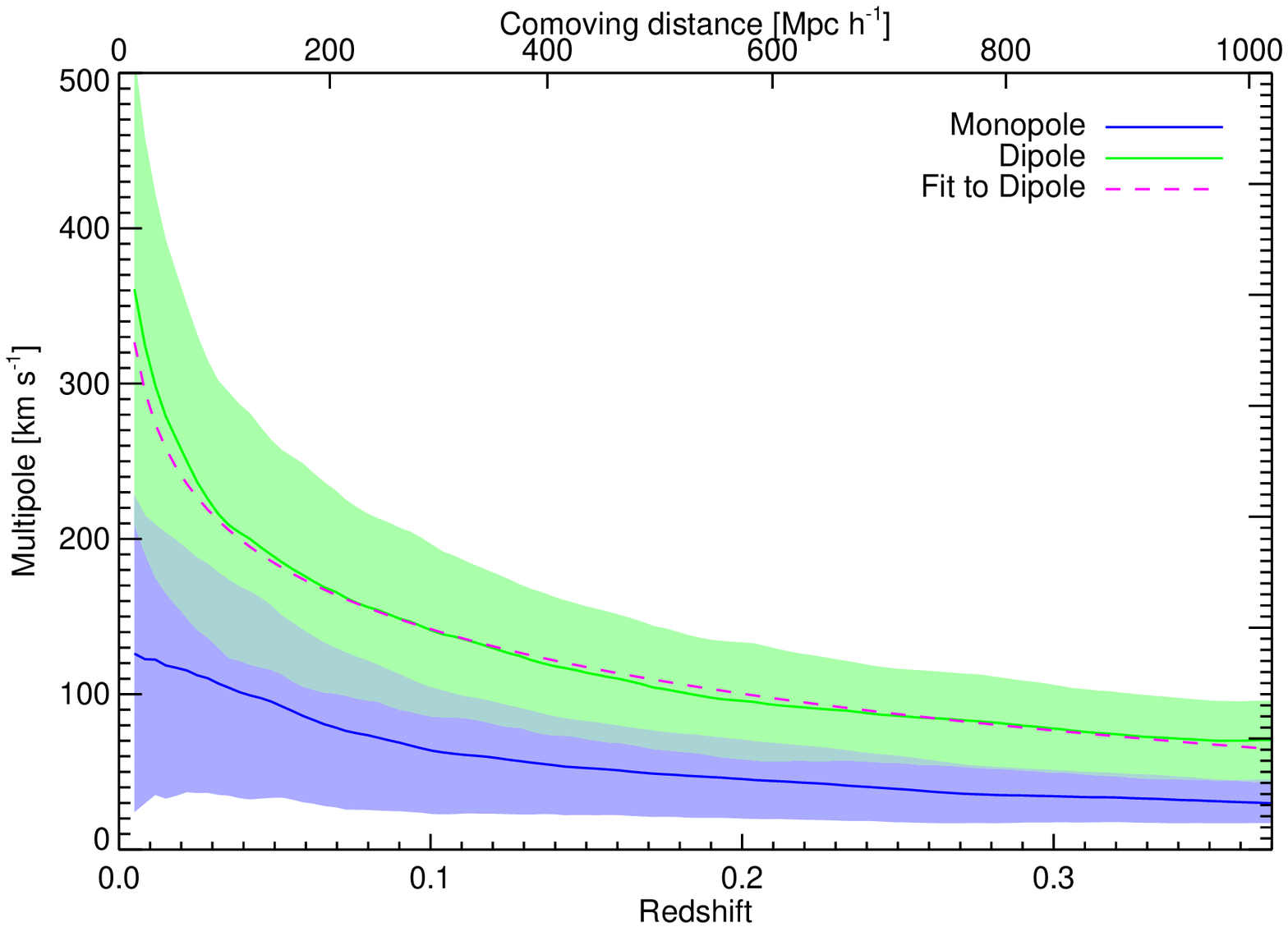}
\includegraphics[width=84mm]{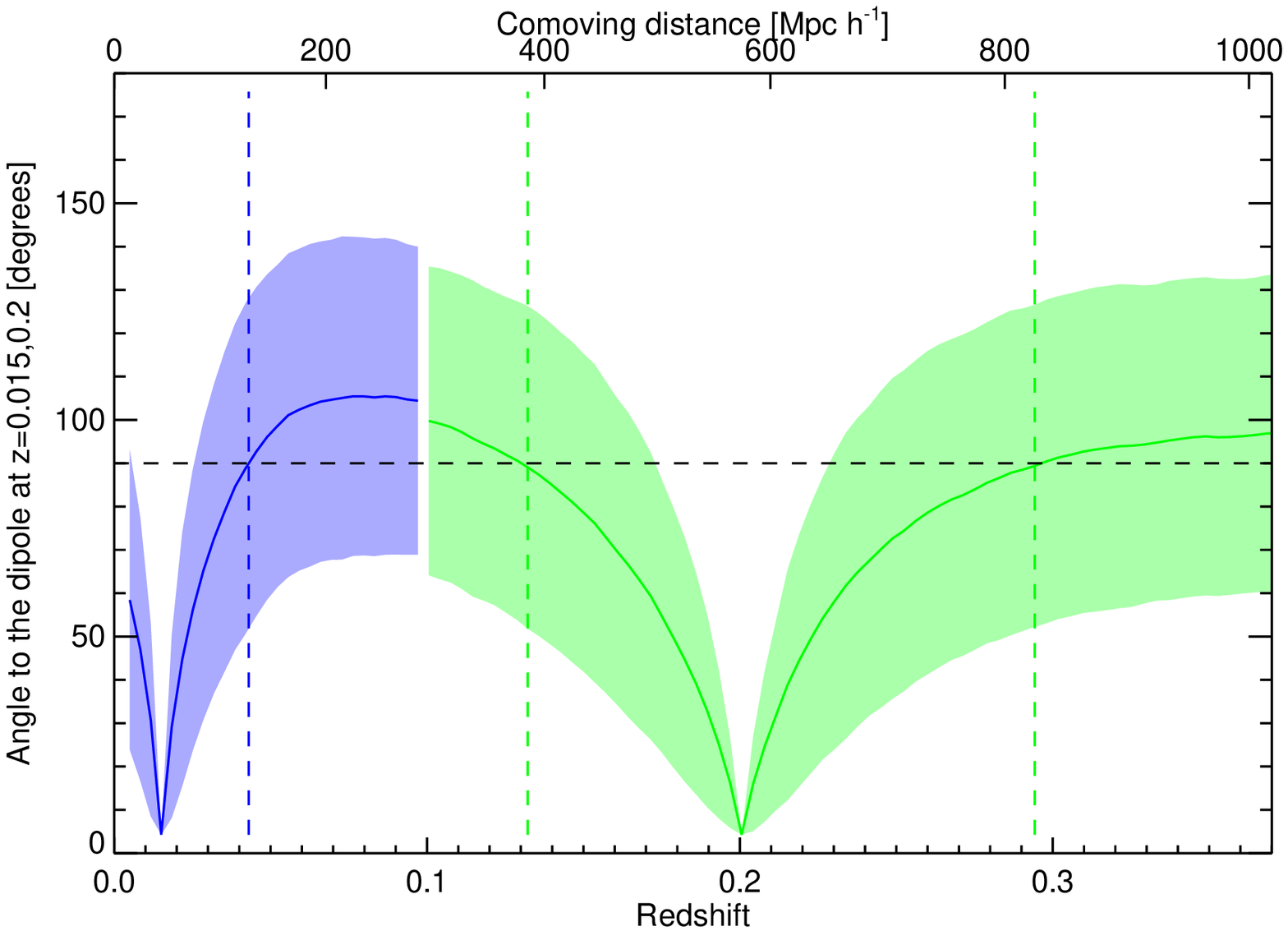}
\caption{Left: The average monopole (blue) and dipole (green) in the velocity field as
extracted from a large N-body simulation with best-fit WMAP5
cosmological parameters. The shaded areas indicate the cosmic
variance. The dashed line is the best-fit model, 
Eq.~(\ref{eq:dip}).   Right: To what distance can the direction of the local dipole be extrapolated?  
This plot shows the relative direction of the dipole as measured by observers at two
reference redshifts $z_0=0.015,0.2$ (blue and green respectively).  When the change reaches  $90^{\circ}$ (dashed lines), which occurs at $z_{90}$=0.043 and $z_{90}$=\{0.132,0.294\}, the dipole to a shell at that radius bears no correlation with the dipole in shells very close to the observer.  
The shaded area is the cosmic variance. As expected, the size of the region in which 
the directions of the dipole are aligned increases with redshift, from 
$\Delta z_{90}\sim$0.03 to $\Delta z_{90}\sim$0.08 for the two cases shown here. }
\label{fig:monodip}\label{fig:theta}
\end{figure*}

\section{Appendix C: Modelling the size of density fluctuations}

Once we have derived a model (such as $\Lambda$CDM) from the observational data we can ask whether the density fluctuations predicted in this model are consistent with those observed.  We can also address whether our treatment of the dipole is justified and whether it is likely that an undiagnosed monopole term could be biasing our results.  This checks the internal consistency of the model as well as testing for biases our assumptions may impose on our results. 

To this end we performed a large scale dark matter N-body simulation ($L_{\textrm{box}}$=2048 Mpc $h^{-1}$, $N_{\textrm{part}}$=$1024^3$, $z_{\textrm{start}}$=49) using the Gadget2 code \citep{Gadget2} with best-fit WMAP5 cosmological parameters
\{$\Omega_m,\Omega_\Lambda,h,w,n_s,\sigma_8$\} = \{0.2792, 0.7208, 0.701, -1, 0.96, 0.817\}, \citep{komatsu09}.
Using 2000 observers placed at random, but weighted by mass, we
calculated the average magnitude of the monopole and the dipole, together with the
cosmic variance of each (see Fig.~\ref{fig:monodip}).  
We define the magnitude of the monopole to be 
$\sigma(M) = \sqrt{{\pi \over 2}} \langle |M| \rangle$, where $\sigma(M)$ is the root-mean-square of the signed monopole, and the signed monopole is simply the mean velocity (either towards or away) of matter in a shell of a particular distance. 

While the analysis is done
at redshift zero using a single data snapshot, the velocities have been corrected
using linear theory, so the velocities in Fig.~\ref{fig:monodip} are in the lightcone. The
correction from translating the velocities to the lightcone is minor, at maximum 1\%.

In order to calculate the mean size of the dipole and how it varies with distance (redshift) we sliced the resulting simulation into shells with a range of radii between 10 \hinvMpc~$ < r < $~1000 \hinvMpc\, around each of the random observers.  We then measured the mean motion of these shells to calculate the monopole and dipole the central observer would see for sources at that distance.  

We have done the analysis both with shells with a thickness of 10 \hinvMpc\,  and of 1 \hinvMpc, and confirmed
that the results do not depend on the shell thickness, except at the very lowest redshifts
where the thickness becomes comparable to the radius of the shell. The magnitude of the dipole is
reasonably well described by the simple model
\begin{align}\label{eq:dip}
v_d&= (507\pm51) - (65\pm8) \ln\left({R_0\chi\over 1 \textrm{Mpc}\,h^{-1}} \right)
           \,\,\textrm{km s}^{-1}, \\ \nonumber
&  = (-8\pm12) - (63\pm7) \ln(z) \,\,\textrm{km s}^{-1}\,,
\end{align}
where the differences in the fits using either comoving distance ($R_0\chi$) or redshifts are due to the slightly nonlinear conversion between the two at larger distances. 

The mean magnitude of the dipole at low redshifts ($z\sim0.01$) is approximately 300 $\pm$ 100 \kms.   
This is consistent with the average random peculiar velocity uncertainty we assume for supernovae.

The mean absolute magnitude of the monopole, which is also plotted in Fig.~\ref{fig:monodip}, is smaller than the dipole but still significant, on the order of 100~\kms. 
We investigate in Section~\ref{sect:monopole} the impact this mean monopole would have on our cosmological inferences. 

The direction of the dipole of the local velocity field is only known at very
low redshifts. To test how well this knowledge can be extrapolated to higher redshifts, 
Fig.~\ref{fig:theta} shows how the direction of the dipole in the simulation changes as a function of redshift.  
The direction is measured with respect to the direction of the dipole at two reference redshifts $z=0.015, 0.2$.  These redshifts correspond respectively to the redshift of the currently available local dipole measurement, and a redshift representative of the SDSS supernovae. 

A na\"{i}ve expectation would be that the average local dipole should decrease with redshift until we reach sources that are too distant to share any significant common source of gravitational attraction with us.  At that point the sources should be on-average at rest with respect to the CMB, and therefore our dipole direction with respect to those sources, if we do not correct for our local velocity, should be simply the direction of the CMB dipole. When we do correct for the local velocity (as done in Figure \ref{fig:theta}), while the amplitude of the local dipole decreases, the direction still changes at higher redshifts, and
it only makes sense to extrapolate the currently known dipole direction out to $z\approx0.045$.

This result is interesting seen in light of observational results pointing towards a coherent
dipole direction out to at least 300 \hinvMpc\, \citep[i.e. $z\sim0.1$,][]{kashlinsky08}, since not only the magnitude of the observed dipole velocity (1600$\pm$500\kms\, at $z\sim0.03$ and $850\pm250$\kms\, at $z\sim0.1$) but also the constancy of the the direction of the dipole is surprising ($\sim3\sigma$ deviation) when interpreted in the framework of the $\Lambda$CDM cosmology.




\bibliographystyle{hapj_tam}
\bibliography{bib09}

\begin{thebibliography}{77}
\expandafter\ifx\csname natexlab\endcsname\relax\def\natexlab#1{#1}\fi

\bibitem[{{Abate} \& {Lahav}(2008)}]{abate08}
{Abate}, A., \& {Lahav}, O. 2008, \mnras, 389, L47

\bibitem[{{Alexander} {et~al.}(2009){Alexander}, {Biswas}, {Notari}, \&
  {Vaid}}]{alexander09}
{Alexander}, S., {Biswas}, T., {Notari}, A., \& {Vaid}, D. 2009, Journal of
  Cosmology and Astro-Particle Physics, 9, 25

\bibitem[{{Alnes} \& {Amarzguioui}(2006)}]{alnes06b}
{Alnes}, H., \& {Amarzguioui}, M. 2006, \prd, 74, 103520

\bibitem[{{Alnes} {et~al.}(2006){Alnes}, {Amarzguioui}, \&
  {Gr{\o}n}}]{alnes06a}
{Alnes}, H., {Amarzguioui}, M., \& {Gr{\o}n}, {\O}. 2006, \prd, 73, 083519

\bibitem[{{Astier} {et~al.}(2006){Astier}, {Guy}, {Regnault}, {Pain},
  {Aubourg}, {Balam}, {Basa}, {Carlberg}, {Fabbro}, {Fouchez}, {Hook},
  {Howell}, {Lafoux}, {Neill}, {Palanque-Delabrouille}, {Perrett}, {Pritchet},
  {Rich}, {Sullivan}, {Taillet}, {Aldering}, {Antilogus}, {Arsenijevic},
  {Balland}, {Baumont}, {Bronder}, {Courtois}, {Ellis}, {Filiol}, {Gon{\c
  c}alves}, {Goobar}, {Guide}, {Hardin}, {Lusset}, {Lidman}, {McMahon},
  {Mouchet}, {Mourao}, {Perlmutter}, {Ripoche}, {Tao}, \& {Walton}}]{astier06}
{Astier}, P. {et~al.} 2006, \aap, 447, 31

\bibitem[{{Bennett} {et~al.}(2003){Bennett}, {Halpern}, {Hinshaw}, {Jarosik},
  {Kogut}, {Limon}, {Meyer}, {Page}, {Spergel}, {Tucker}, {Wollack}, {Wright},
  {Barnes}, {Greason}, {Hill}, {Komatsu}, {Nolta}, {Odegard}, {Peiris},
  {Verde}, \& {Weiland}}]{bennett03}
{Bennett}, C.~L. {et~al.} 2003, \apjs, 148, 1

\bibitem[{{Blake} {et~al.}(2011{\natexlab{a}}){Blake}, {Brough}, {Colless},
  {Contreras}, {Couch}, {Croom}, {Davis}, {Drinkwater}, {Forster}, {Gilbank},
  {Gladders}, {Glazebrook}, {Jelliffe}, {Jurek}, {Li}, {Madore}, {Martin},
  {Pimbblet}, {Poole}, {Pracy}, {Sharp}, {Wisnioski}, {Woods}, {Wyder}, \&
  {Yee}}]{blake11a}
{Blake}, C. {et~al.} 2011{\natexlab{a}}, ArXiv e-prints

\bibitem[{{Blake} {et~al.}(2011{\natexlab{b}}){Blake}, {Davis}, {Poole},
  {Parkinson}, {Brough}, {Colless}, {Contreras}, {Couch}, {Croom},
  {Drinkwater}, {Forster}, {Gilbank}, {Gladders}, {Glazebrook}, {Jelliffe},
  {Jurek}, {Li}, {Madore}, {Martin}, {Pimbblet}, {Pracy}, {Sharp}, {Wisnioski},
  {Woods}, {Wyder}, \& {Yee}}]{blake11b}
------. 2011{\natexlab{b}}, ArXiv e-prints

\bibitem[{{Blomqvist} \& {M{\"o}rtsell}(2010)}]{blomqvist10}
{Blomqvist}, M., \& {M{\"o}rtsell}, E. 2010, Journal of Cosmology and
  Astro-Particle Physics, 5, 6

\bibitem[{{Blondin} {et~al.}(2011){Blondin}, {Mandel}, \&
  {Kirshner}}]{blondin10}
{Blondin}, S., {Mandel}, K.~S., \& {Kirshner}, R.~P. 2011, \aap, 526, A81+

\bibitem[{{Bonvin} {et~al.}(2006{\natexlab{a}}){Bonvin}, {Durrer}, \&
  {Gasparini}}]{bonvin06flucts}
{Bonvin}, C., {Durrer}, R., \& {Gasparini}, M.~A. 2006{\natexlab{a}}, \prd, 73,
  023523

\bibitem[{{Bonvin} {et~al.}(2006{\natexlab{b}}){Bonvin}, {Durrer}, \&
  {Kunz}}]{bonvin06dipole}
{Bonvin}, C., {Durrer}, R., \& {Kunz}, M. 2006{\natexlab{b}}, Physical Review
  Letters, 96, 191302

\bibitem[{{Caldwell} \& {Stebbins}(2008)}]{caldwell08}
{Caldwell}, R.~R., \& {Stebbins}, A. 2008, Physical Review Letters, 100, 191302

\bibitem[{{Clifton} {et~al.}(2008){Clifton}, {Ferreira}, \& {Land}}]{clifton08}
{Clifton}, T., {Ferreira}, P.~G., \& {Land}, K. 2008, Physical Review Letters,
  101, 131302

\bibitem[{{Conley} {et~al.}(2007){Conley}, {Carlberg}, {Guy}, {Howell}, {Jha},
  {Riess}, \& {Sullivan}}]{conley07}
{Conley}, A., {Carlberg}, R.~G., {Guy}, J., {Howell}, D.~A., {Jha}, S.,
  {Riess}, A.~G., \& {Sullivan}, M. 2007, \apjl, 664, L13

\bibitem[{{Conley} {et~al.}(2006){Conley}, {Goldhaber}, {Wang}, {Aldering},
  {Amanullah}, {Commins}, {Fadeyev}, {Folatelli}, {Garavini}, {Gibbons},
  {Goobar}, {Groom}, {Hook}, {Howell}, {Kim}, {Knop}, {Kowalski}, {Kuznetsova},
  {Lidman}, {Nobili}, {Nugent}, {Pain}, {Perlmutter}, {Smith}, {Spadafora},
  {Stanishev}, {Strovink}, {Thomas}, \& {Wood-Vasey}}]{conley06}
{Conley}, A. {et~al.} 2006, \apj, 644, 1

\bibitem[{{Cooray} \& {Caldwell}(2006)}]{cooray06}
{Cooray}, A., \& {Caldwell}, R.~R. 2006, \prd, 73, 103002

\bibitem[{{Davis} \& {Lineweaver}(2004)}]{davis04}
{Davis}, T.~M., \& {Lineweaver}, C.~H. 2004, Publications of the Astronomical
  Society of Australia, 21, 97

\bibitem[{{Davis} \& {Lineweaver}(2005)}]{davis05}
------. 2005, Scientific American, 292, 36

\bibitem[{{Davis} {et~al.}(2007){Davis}, {M{\"o}rtsell}, {Sollerman}, {Becker},
  {Blondin}, {Challis}, {Clocchiatti}, {Filippenko}, {Foley}, {Garnavich},
  {Jha}, {Krisciunas}, {Kirshner}, {Leibundgut}, {Li}, {Matheson}, {Miknaitis},
  {Pignata}, {Rest}, {Riess}, {Schmidt}, {Smith}, {Spyromilio}, {Stubbs},
  {Suntzeff}, {Tonry}, {Wood-Vasey}, \& {Zenteno}}]{davis07}
{Davis}, T.~M. {et~al.} 2007, \apj, 666, 716

\bibitem[{{Eisenstein} {et~al.}(2005){Eisenstein}, {Zehavi}, {Hogg},
  {Scoccimarro}, {Blanton}, {Nichol}, {Scranton}, {Seo}, {Tegmark}, {Zheng},
  {Anderson}, {Annis}, {Bahcall}, {Brinkmann}, {Burles}, {Castander},
  {Connolly}, {Csabai}, {Doi}, {Fukugita}, {Frieman}, {Glazebrook}, {Gunn},
  {Hendry}, {Hennessy}, {Ivezi{\'c}}, {Kent}, {Knapp}, {Lin}, {Loh}, {Lupton},
  {Margon}, {McKay}, {Meiksin}, {Munn}, {Pope}, {Richmond}, {Schlegel},
  {Schneider}, {Shimasaku}, {Stoughton}, {Strauss}, {SubbaRao}, {Szalay},
  {Szapudi}, {Tucker}, {Yanny}, \& {York}}]{eisenstein05}
{Eisenstein}, D.~J. {et~al.} 2005, \apj, 633, 560

\bibitem[{{Enqvist}(2008)}]{enqvist08}
{Enqvist}, K. 2008, General Relativity and Gravitation, 40, 451

\bibitem[{{Enqvist} \& {Mattsson}(2007)}]{enqvist07}
{Enqvist}, K., \& {Mattsson}, T. 2007, Journal of Cosmology and Astro-Particle
  Physics, 2, 19

\bibitem[{{Erdo{\u g}du} {et~al.}(2006){Erdo{\u g}du}, {Huchra}, {Lahav},
  {Colless}, {Cutri}, {Falco}, {George}, {Jarrett}, {Jones}, {Kochanek},
  {Macri}, {Mader}, {Martimbeau}, {Pahre}, {Parker}, {Rassat}, \&
  {Saunders}}]{erdogdu06}
{Erdo{\u g}du}, P. {et~al.} 2006, \mnras, 368, 1515

\bibitem[{{Folatelli} {et~al.}(2010){Folatelli}, {Phillips}, {Burns},
  {Contreras}, {Hamuy}, {Freedman}, {Persson}, {Stritzinger}, {Suntzeff},
  {Krisciunas}, {Boldt}, {Gonz{\'a}lez}, {Krzeminski}, {Morrell}, {Roth},
  {Salgado}, {Madore}, {Murphy}, {Wyatt}, {Li}, {Filippenko}, \&
  {Miller}}]{folatelli10}
{Folatelli}, G. {et~al.} 2010, \aj, 139, 120

\bibitem[{{Freedman} {et~al.}(2009){Freedman}, {Burns}, {Phillips}, {Wyatt},
  {Persson}, {Madore}, {Contreras}, {Folatelli}, {Gonzalez}, {Hamuy}, {Hsiao},
  {Kelson}, {Morrell}, {Murphy}, {Roth}, {Stritzinger}, {Sturch}, {Suntzeff},
  {Astier}, {Balland}, {Bassett}, {Boldt}, {Carlberg}, {Conley}, {Frieman},
  {Garnavich}, {Guy}, {Hardin}, {Howell}, {Kessler}, {Lampeitl}, {Marriner},
  {Pain}, {Perrett}, {Regnault}, {Riess}, {Sako}, {Schneider}, {Sullivan}, \&
  {Wood-Vasey}}]{freedman09}
{Freedman}, W.~L. {et~al.} 2009, \apj, 704, 1036

\bibitem[{{Frieman} {et~al.}(2008){Frieman}, {Bassett}, {Becker}, {Choi},
  {Cinabro}, {DeJongh}, {Depoy}, {Dilday}, {Doi}, {Garnavich}, {Hogan},
  {Holtzman}, {Im}, {Jha}, {Kessler}, {Konishi}, {Lampeitl}, {Marriner},
  {Marshall}, {McGinnis}, {Miknaitis}, {Nichol}, {Prieto}, {Riess}, {Richmond},
  {Romani}, {Sako}, {Schneider}, {Smith}, {Takanashi}, {Tokita}, {van der
  Heyden}, {Yasuda}, {Zheng}, {Adelman-McCarthy}, {Annis}, {Assef},
  {Barentine}, {Bender}, {Blandford}, {Boroski}, {Bremer}, {Brewington},
  {Collins}, {Crotts}, {Dembicky}, {Eastman}, {Edge}, {Edmondson}, {Elson},
  {Eyler}, {Filippenko}, {Foley}, {Frank}, {Goobar}, {Gueth}, {Gunn},
  {Harvanek}, {Hopp}, {Ihara}, {Ivezi{\'c}}, {Kahn}, {Kaplan}, {Kent},
  {Ketzeback}, {Kleinman}, {Kollatschny}, {Kron}, {Krzesi{\'n}ski}, {Lamenti},
  {Leloudas}, {Lin}, {Long}, {Lucey}, {Lupton}, {Malanushenko}, {Malanushenko},
  {McMillan}, {Mendez}, {Morgan}, {Morokuma}, {Nitta}, {Ostman}, {Pan},
  {Rockosi}, {Romer}, {Ruiz-Lapuente}, {Saurage}, {Schlesinger}, {Snedden},
  {Sollerman}, {Stoughton}, {Stritzinger}, {Subba Rao}, {Tucker}, {Vaisanen},
  {Watson}, {Watters}, {Wheeler}, {Yanny}, \& {York}}]{frieman08}
{Frieman}, J.~A. {et~al.} 2008, \aj, 135, 338

\bibitem[{{Garcia-Bellido} \& {Haugb{\o}lle}(2008)}]{garcia-bellido08}
{Garcia-Bellido}, J., \& {Haugb{\o}lle}, T. 2008, Journal of Cosmology and
  Astro-Particle Physics, 4, 3

\bibitem[{{Garc{\'{\i}}a-Bellido} \&
  {Haugb{\o}lle}(2008)}]{garcia-bellido08-kz}
{Garc{\'{\i}}a-Bellido}, J., \& {Haugb{\o}lle}, T. 2008, Journal of Cosmology
  and Astro-Particle Physics, 9, 16

\bibitem[{{Garc{\'{\i}}a-Bellido} \& {Haugb{\o}lle}(2009)}]{garcia-bellido09}
------. 2009, Journal of Cosmology and Astro-Particle Physics, 9, 28

\bibitem[{{Geller} {et~al.}(1997){Geller}, {Kurtz}, {Wegner}, {Thorstensen},
  {Fabricant}, {Marzke}, {Huchra}, {Schild}, \& {Falco}}]{geller97}
{Geller}, M.~J. {et~al.} 1997, \aj, 114, 2205

\bibitem[{{Gordon} {et~al.}(2007){Gordon}, {Land}, \& {Slosar}}]{gordon07}
{Gordon}, C., {Land}, K., \& {Slosar}, A. 2007, Physical Review Letters, 99,
  081301

\bibitem[{{Gordon} {et~al.}(2008){Gordon}, {Land}, \& {Slosar}}]{gordon08}
------. 2008, \mnras, 387, 371

\bibitem[{{Gorski}(1988)}]{gorski88}
{Gorski}, K. 1988, \apjl, 332, L7

\bibitem[{{G{\'o}rski} {et~al.}(2005){G{\'o}rski}, {Hivon}, {Banday},
  {Wandelt}, {Hansen}, {Reinecke}, \& {Bartelmann}}]{gorski05}
{G{\'o}rski}, K.~M., {Hivon}, E., {Banday}, A.~J., {Wandelt}, B.~D., {Hansen},
  F.~K., {Reinecke}, M., \& {Bartelmann}, M. 2005, \apj, 622, 759

\bibitem[{Gottl\"ober {et~al.}(2010)Gottl\"ober, Hoffman, \&
  Yepes}]{gottloeber10}
Gottl\"ober, S., Hoffman, Y., \& Yepes, G. 2010, in High Performance Computing
  in Science and Engineering, Garching/Munich 2009, ed. S.~Wagner,
  M.~Steinmetz, A.~Bode, \& M.~M. MŸller (Springer Berlin Heidelberg), 309--322

\bibitem[{{Hamuy} {et~al.}(1996){Hamuy}, {Phillips}, {Suntzeff}, {Schommer},
  {Maza}, {Antezan}, {Wischnjewsky}, {Valladares}, {Muena}, {Gonzales},
  {Aviles}, {Wells}, {Smith}, {Navarrete}, {Covarrubias}, {Williger}, {Walker},
  {Layden}, {Elias}, {Baldwin}, {Hernandez}, {Tirado}, {Ugarte}, {Elston},
  {Saavedra}, {Barrientos}, {Costa}, {Lira}, {Ruiz}, {Anguita}, {Gomez},
  {Ortiz}, {della Valle}, {Danziger}, {Storm}, {Kim}, {Bailyn}, {Rubenstein},
  {Tucker}, {Cersosimo}, {Mendez}, {Siciliano}, {Sherry}, {Chaboyer},
  {Koopmann}, {Geisler}, {Sarajedini}, {Dey}, {Tyson}, {Rich}, {Gal},
  {Lamontagne}, {Caldwell}, {Guhathakurta}, {Phillips}, {Szkody}, {Prosser},
  {Ho}, {McMahan}, {Baggley}, {Cheng}, {Havlen}, {Wakamatsu}, {Janes},
  {Malkan}, {Baganoff}, {Seitzer}, {Shara}, {Sturch}, {Hesser}, {Hartig},
  {Hughes}, {Welch}, {Williams}, {Ferguson}, {Francis}, {French}, {Bolte},
  {Roth}, {Odewahn}, {Howell}, \& {Krzeminski}}]{hamuy96}
{Hamuy}, M. {et~al.} 1996, \aj, 112, 2408

\bibitem[{{Hannestad} {et~al.}(2008){Hannestad}, {Haugb{\o}lle}, \&
  {Thomsen}}]{hannestad08}
{Hannestad}, S., {Haugb{\o}lle}, T., \& {Thomsen}, B. 2008, Journal of
  Cosmology and Astro-Particle Physics, 2, 22

\bibitem[{{Haugb{\o}lle} {et~al.}(2007){Haugb{\o}lle}, {Hannestad}, {Thomsen},
  {Fynbo}, {Sollerman}, \& {Jha}}]{haugbolle07}
{Haugb{\o}lle}, T., {Hannestad}, S., {Thomsen}, B., {Fynbo}, J., {Sollerman},
  J., \& {Jha}, S. 2007, \apj, 661, 650

\bibitem[{{Hern{\'a}ndez-Monteagudo} {et~al.}(2006){Hern{\'a}ndez-Monteagudo},
  {Verde}, {Jimenez}, \& {Spergel}}]{hernandez06}
{Hern{\'a}ndez-Monteagudo}, C., {Verde}, L., {Jimenez}, R., \& {Spergel}, D.~N.
  2006, \apj, 643, 598

\bibitem[{{Hicken} {et~al.}(2009){Hicken}, {Wood-Vasey}, {Blondin}, {Challis},
  {Jha}, {Kelly}, {Rest}, \& {Kirshner}}]{hicken09}
{Hicken}, M., {Wood-Vasey}, W.~M., {Blondin}, S., {Challis}, P., {Jha}, S.,
  {Kelly}, P.~L., {Rest}, A., \& {Kirshner}, R.~P. 2009, \apj, 700, 1097

\bibitem[{{Holtzman} {et~al.}(2008){Holtzman}, {Marriner}, {Kessler}, {Sako},
  {Dilday}, {Frieman}, {Schneider}, {Bassett}, {Becker}, {Cinabro}, {DeJongh},
  {Depoy}, {Doi}, {Garnavich}, {Hogan}, {Jha}, {Konishi}, {Lampeitl},
  {Marshall}, {McGinnis}, {Miknaitis}, {Nichol}, {Prieto}, {Riess}, {Richmond},
  {Romani}, {Smith}, {Takanashi}, {Tokita}, {van der Heyden}, {Yasuda}, \&
  {Zheng}}]{holtzman08}
{Holtzman}, J.~A. {et~al.} 2008, \aj, 136, 2306

\bibitem[{{Hui} \& {Greene}(2006)}]{hui06}
{Hui}, L., \& {Greene}, P.~B. 2006, \prd, 73, 123526

\bibitem[{{Jha} {et~al.}(2006){Jha}, {Kirshner}, {Challis}, {Garnavich},
  {Matheson}, {Soderberg}, {Graves}, {Hicken}, {Alves}, {Arce}, {Balog},
  {Barmby}, {Barton}, {Berlind}, {Bragg}, {Brice{\~n}o}, {Brown}, {Buckley},
  {Caldwell}, {Calkins}, {Carter}, {Concannon}, {Donnelly}, {Eriksen},
  {Fabricant}, {Falco}, {Fiore}, {Garcia}, {G{\'o}mez}, {Grogin}, {Groner},
  {Groot}, {Haisch}, {Hartmann}, {Hergenrother}, {Holman}, {Huchra},
  {Jayawardhana}, {Jerius}, {Kannappan}, {Kim}, {Kleyna}, {Kochanek},
  {Koranyi}, {Krockenberger}, {Lada}, {Luhman}, {Luu}, {Macri}, {Mader},
  {Mahdavi}, {Marengo}, {Marsden}, {McLeod}, {McNamara}, {Megeath}, {Moraru},
  {Mossman}, {Muench}, {Mu{\~n}oz}, {Muzerolle}, {Naranjo}, {Nelson-Patel},
  {Pahre}, {Patten}, {Peters}, {Peters}, {Raymond}, {Rines}, {Schild},
  {Sobczak}, {Spahr}, {Stauffer}, {Stefanik}, {Szentgyorgyi}, {Tollestrup},
  {V{\"a}is{\"a}nen}, {Vikhlinin}, {Wang}, {Willner}, {Wolk}, {Zajac}, {Zhao},
  \& {Stanek}}]{jha06}
{Jha}, S. {et~al.} 2006, \aj, 131, 527

\bibitem[{{Jha} {et~al.}(2007){Jha}, {Riess}, \& {Kirshner}}]{jha07}
{Jha}, S., {Riess}, A.~G., \& {Kirshner}, R.~P. 2007, \apj, 659, 122

\bibitem[{{Kashlinsky} {et~al.}(2008){Kashlinsky}, {Atrio-Barandela},
  {Kocevski}, \& {Ebeling}}]{kashlinsky08}
{Kashlinsky}, A., {Atrio-Barandela}, F., {Kocevski}, D., \& {Ebeling}, H. 2008,
  \apjl, 686, L49

\bibitem[{{Kashlinsky} {et~al.}(2009){Kashlinsky}, {Atrio-Barandela},
  {Kocevski}, \& {Ebeling}}]{kashlinsky09}
------. 2009, \apj, 691, 1479

\bibitem[{{Kessler} {et~al.}(2009){Kessler}, {Becker}, {Cinabro}, {Vanderplas},
  {Frieman}, {Marriner}, {Davis}, {Dilday}, {Holtzman}, {Jha}, {Lampeitl},
  {Sako}, {Smith}, {Zheng}, {Nichol}, {Bassett}, {Bender}, {Depoy}, {Doi},
  {Elson}, {Filippenko}, {Foley}, {Garnavich}, {Hopp}, {Ihara}, {Ketzeback},
  {Kollatschny}, {Konishi}, {Marshall}, {Mc Millan}, {Miknaitis}, {Morokuma},
  {M{\"o}rtsell}, {Pan}, {Prieto}, {Richmond}, {Riess}, {Romani}, {Schneider},
  {Sollerman}, {Takanashi}, {Tokita}, {van der Heyden}, {Wheeler}, {Yasuda}, \&
  {York}}]{kessler09}
{Kessler}, R. {et~al.} 2009, \apjs, 185, 32

\bibitem[{{Kogut} {et~al.}(1993){Kogut}, {Lineweaver}, {Smoot}, {Bennett},
  {Banday}, {Boggess}, {Cheng}, {de Amici}, {Fixsen}, {Hinshaw}, {Jackson},
  {Janssen}, {Keegstra}, {Loewenstein}, {Lubin}, {Mather}, {Tenorio}, {Weiss},
  {Wilkinson}, \& {Wright}}]{kogut93}
{Kogut}, A. {et~al.} 1993, \apj, 419, 1

\bibitem[{{Komatsu} {et~al.}(2009){Komatsu}, {Dunkley}, {Nolta}, {Bennett},
  {Gold}, {Hinshaw}, {Jarosik}, {Larson}, {Limon}, {Page}, {Spergel},
  {Halpern}, {Hill}, {Kogut}, {Meyer}, {Tucker}, {Weiland}, {Wollack}, \&
  {Wright}}]{komatsu09}
{Komatsu}, E. {et~al.} 2009, \apjs, 180, 330

\bibitem[{{Komatsu} {et~al.}(2011){Komatsu}, {Smith}, {Dunkley}, {Bennett},
  {Gold}, {Hinshaw}, {Jarosik}, {Larson}, {Nolta}, {Page}, {Spergel},
  {Halpern}, {Hill}, {Kogut}, {Limon}, {Meyer}, {Odegard}, {Tucker}, {Weiland},
  {Wollack}, \& {Wright}}]{komatsu11}
------. 2011, \apjs, 192, 18

\bibitem[{{Kowalski} {et~al.}(2008){Kowalski}, {Rubin}, {Aldering},
  {Agostinho}, {Amadon}, {Amanullah}, {Balland}, {Barbary}, {Blanc}, {Challis},
  {Conley}, {Connolly}, {Covarrubias}, {Dawson}, {Deustua}, {Ellis}, {Fabbro},
  {Fadeyev}, {Fan}, {Farris}, {Folatelli}, {Frye}, {Garavini}, {Gates},
  {Germany}, {Goldhaber}, {Goldman}, {Goobar}, {Groom}, {Haissinski}, {Hardin},
  {Hook}, {Kent}, {Kim}, {Knop}, {Lidman}, {Linder}, {Mendez}, {Meyers},
  {Miller}, {Moniez}, {Mour{\~a}o}, {Newberg}, {Nobili}, {Nugent}, {Pain},
  {Perdereau}, {Perlmutter}, {Phillips}, {Prasad}, {Quimby}, {Regnault},
  {Rich}, {Rubenstein}, {Ruiz-Lapuente}, {Santos}, {Schaefer}, {Schommer},
  {Smith}, {Soderberg}, {Spadafora}, {Strolger}, {Strovink}, {Suntzeff},
  {Suzuki}, {Thomas}, {Walton}, {Wang}, {Wood-Vasey}, \& {Yun}}]{kowalski08}
{Kowalski}, M. {et~al.} 2008, \apj, 686, 749

\bibitem[{{Lampeitl} {et~al.}(2010){Lampeitl}, {Nichol}, {Seo}, {Giannantonio},
  {Shapiro}, {Bassett}, {Percival}, {Davis}, {Dilday}, {Frieman}, {Garnavich},
  {Sako}, {Smith}, {Sollerman}, {Becker}, {Cinabro}, {Filippenko}, {Foley},
  {Hogan}, {Holtzman}, {Jha}, {Konishi}, {Marriner}, {Richmond}, {Riess},
  {Schneider}, {Stritzinger}, {van der Heyden}, {Vanderplas}, {Wheeler}, \&
  {Zheng}}]{lampeitl10}
{Lampeitl}, H. {et~al.} 2010, \mnras, 401, 2331

\bibitem[{{Marra} \& {P{\"a}{\"a}kk{\"o}nen}(2010)}]{marra10}
{Marra}, V., \& {P{\"a}{\"a}kk{\"o}nen}, M. 2010, Journal of Cosmology and
  Astro-Particle Physics, 12, 21

\bibitem[{{NASA/IPAC Extragalactic Database}(2008)}]{NED}
{NASA/IPAC Extragalactic Database}. 2008, {NED Velocity Correction Calculator}

\bibitem[{{Neill} {et~al.}(2007){Neill}, {Hudson}, \& {Conley}}]{neill07}
{Neill}, J.~D., {Hudson}, M.~J., \& {Conley}, A. 2007, \apjl, 661, L123

\bibitem[{{Page} {et~al.}(2003){Page}, {Nolta}, {Barnes}, {Bennett}, {Halpern},
  {Hinshaw}, {Jarosik}, {Kogut}, {Limon}, {Meyer}, {Peiris}, {Spergel},
  {Tucker}, {Wollack}, \& {Wright}}]{page03}
{Page}, L. {et~al.} 2003, \apjs, 148, 233

\bibitem[{{Percival} {et~al.}(2007){Percival}, {Cole}, {Eisenstein}, {Nichol},
  {Peacock}, {Pope}, \& {Szalay}}]{percival07}
{Percival}, W.~J., {Cole}, S., {Eisenstein}, D.~J., {Nichol}, R.~C., {Peacock},
  J.~A., {Pope}, A.~C., \& {Szalay}, A.~S. 2007, \mnras, 381, 1053

\bibitem[{{Percival} {et~al.}(2010){Percival}, {Reid}, {Eisenstein}, {Bahcall},
  {Budavari}, {Frieman}, {Fukugita}, {Gunn}, {Ivezi{\'c}}, {Knapp}, {Kron},
  {Loveday}, {Lupton}, {McKay}, {Meiksin}, {Nichol}, {Pope}, {Schlegel},
  {Schneider}, {Spergel}, {Stoughton}, {Strauss}, {Szalay}, {Tegmark},
  {Vogeley}, {Weinberg}, {York}, \& {Zehavi}}]{percival10}
{Percival}, W.~J. {et~al.} 2010, \mnras, 401, 2148

\bibitem[{{Perlmutter} {et~al.}(1999){Perlmutter}, {Aldering}, {Goldhaber},
  {Knop}, {Nugent}, {Castro}, {Deustua}, {Fabbro}, {Goobar}, {Groom}, {Hook},
  {Kim}, {Kim}, {Lee}, {Nunes}, {Pain}, {Pennypacker}, {Quimby}, {Lidman},
  {Ellis}, {Irwin}, {McMahon}, {Ruiz-Lapuente}, {Walton}, {Schaefer}, {Boyle},
  {Filippenko}, {Matheson}, {Fruchter}, {Panagia}, {Newberg}, {Couch}, \& {The
  Supernova Cosmology Project}}]{perlmutter99}
{Perlmutter}, S. {et~al.} 1999, \apj, 517, 565

\bibitem[{{Pyne} \& {Birkinshaw}(1996)}]{pyne96}
{Pyne}, T., \& {Birkinshaw}, M. 1996, \apj, 458, 46

\bibitem[{{Riess} {et~al.}(1998){Riess}, {Filippenko}, {Challis},
  {Clocchiatti}, {Diercks}, {Garnavich}, {Gilliland}, {Hogan}, {Jha},
  {Kirshner}, {Leibundgut}, {Phillips}, {Reiss}, {Schmidt}, {Schommer},
  {Smith}, {Spyromilio}, {Stubbs}, {Suntzeff}, \& {Tonry}}]{riess98}
{Riess}, A.~G. {et~al.} 1998, \aj, 116, 1009

\bibitem[{{Riess} {et~al.}(1999){Riess}, {Kirshner}, {Schmidt}, {Jha},
  {Challis}, {Garnavich}, {Esin}, {Carpenter}, {Grashius}, {Schild}, {Berlind},
  {Huchra}, {Prosser}, {Falco}, {Benson}, {Brice{\~n}o}, {Brown}, {Caldwell},
  {dell'Antonio}, {Filippenko}, {Goodman}, {Grogin}, {Groner}, {Hughes},
  {Green}, {Jansen}, {Kleyna}, {Luu}, {Macri}, {McLeod}, {McLeod}, {McNamara},
  {McLean}, {Milone}, {Mohr}, {Moraru}, {Peng}, {Peters}, {Prestwich},
  {Stanek}, {Szentgyorgyi}, \& {Zhao}}]{riess99}
------. 1999, \aj, 117, 707

\bibitem[{{Riess} {et~al.}(2007){Riess}, {Strolger}, {Casertano}, {Ferguson},
  {Mobasher}, {Gold}, {Challis}, {Filippenko}, {Jha}, {Li}, {Tonry}, {Foley},
  {Kirshner}, {Dickinson}, {MacDonald}, {Eisenstein}, {Livio}, {Younger}, {Xu},
  {Dahl{\'e}n}, \& {Stern}}]{riess07}
------. 2007, \apj, 659, 98

\bibitem[{{Riess} {et~al.}(2004){Riess}, {Strolger}, {Tonry}, {Casertano},
  {Ferguson}, {Mobasher}, {Challis}, {Filippenko}, {Jha}, {Li}, {Chornock},
  {Kirshner}, {Leibundgut}, {Dickinson}, {Livio}, {Giavalisco}, {Steidel},
  {Ben{\'{\i}}tez}, \& {Tsvetanov}}]{riess04}
------. 2004, \apj, 607, 665

\bibitem[{{Sako} {et~al.}(2008){Sako}, {Bassett}, {Becker}, {Cinabro},
  {DeJongh}, {Depoy}, {Dilday}, {Doi}, {Frieman}, {Garnavich}, {Hogan},
  {Holtzman}, {Jha}, {Kessler}, {Konishi}, {Lampeitl}, {Marriner}, {Miknaitis},
  {Nichol}, {Prieto}, {Riess}, {Richmond}, {Romani}, {Schneider}, {Smith},
  {Subba Rao}, {Takanashi}, {Tokita}, {van der Heyden}, {Yasuda}, {Zheng},
  {Barentine}, {Brewington}, {Choi}, {Dembicky}, {Harnavek}, {Ihara}, {Im},
  {Ketzeback}, {Kleinman}, {Krzesi{\'n}ski}, {Long}, {Malanushenko},
  {Malanushenko}, {McMillan}, {Morokuma}, {Nitta}, {Pan}, {Saurage}, \&
  {Snedden}}]{sako08}
{Sako}, M. {et~al.} 2008, \aj, 135, 348

\bibitem[{{Sasaki}(1987)}]{sasaki87}
{Sasaki}, M. 1987, \mnras, 228, 653

\bibitem[{{Sinclair} {et~al.}(2010){Sinclair}, {Davis}, \&
  {Haugb{\o}lle}}]{sinclair10}
{Sinclair}, B., {Davis}, T.~M., \& {Haugb{\o}lle}, T. 2010, \apj, 718, 1445

\bibitem[{{Spergel} {et~al.}(2007){Spergel}, {Bean}, {Dor{\'e}}, {Nolta},
  {Bennett}, {Dunkley}, {Hinshaw}, {Jarosik}, {Komatsu}, {Page}, {Peiris},
  {Verde}, {Halpern}, {Hill}, {Kogut}, {Limon}, {Meyer}, {Odegard}, {Tucker},
  {Weiland}, {Wollack}, \& {Wright}}]{spergel07}
{Spergel}, D.~N. {et~al.} 2007, \apjs, 170, 377

\bibitem[{{Springel}(2005)}]{Gadget2}
{Springel}, V. 2005, \mnras, 364, 1105

\bibitem[{{Tegmark} {et~al.}(2006){Tegmark}, {Eisenstein}, {Strauss},
  {Weinberg}, {Blanton}, {Frieman}, {Fukugita}, {Gunn}, {Hamilton}, {Knapp},
  {Nichol}, {Ostriker}, {Padmanabhan}, {Percival}, {Schlegel}, {Schneider},
  {Scoccimarro}, {Seljak}, {Seo}, {Swanson}, {Szalay}, {Vogeley}, {Yoo},
  {Zehavi}, {Abazajian}, {Anderson}, {Annis}, {Bahcall}, {Bassett}, {Berlind},
  {Brinkmann}, {Budavari}, {Castander}, {Connolly}, {Csabai}, {Doi},
  {Finkbeiner}, {Gillespie}, {Glazebrook}, {Hennessy}, {Hogg}, {Ivezi{\'c}},
  {Jain}, {Johnston}, {Kent}, {Lamb}, {Lee}, {Lin}, {Loveday}, {Lupton},
  {Munn}, {Pan}, {Park}, {Peoples}, {Pier}, {Pope}, {Richmond}, {Rockosi},
  {Scranton}, {Sheth}, {Stebbins}, {Stoughton}, {Szapudi}, {Tucker}, {vanden
  Berk}, {Yanny}, \& {York}}]{tegmark06}
{Tegmark}, M. {et~al.} 2006, \prd, 74, 123507

\bibitem[{{Watkins} {et~al.}(2009){Watkins}, {Feldman}, \&
  {Hudson}}]{watkins09}
{Watkins}, R., {Feldman}, H.~A., \& {Hudson}, M.~J. 2009, \mnras, 392, 743

\bibitem[{{Wood-Vasey} {et~al.}(2007){Wood-Vasey}, {Miknaitis}, {Stubbs},
  {Jha}, {Riess}, {Garnavich}, {Kirshner}, {Aguilera}, {Becker}, {Blackman},
  {Blondin}, {Challis}, {Clocchiatti}, {Conley}, {Covarrubias}, {Davis},
  {Filippenko}, {Foley}, {Garg}, {Hicken}, {Krisciunas}, {Leibundgut}, {Li},
  {Matheson}, {Miceli}, {Narayan}, {Pignata}, {Prieto}, {Rest}, {Salvo},
  {Schmidt}, {Smith}, {Sollerman}, {Spyromilio}, {Tonry}, {Suntzeff}, \&
  {Zenteno}}]{WV07}
{Wood-Vasey}, W.~M. {et~al.} 2007, \apj, 666, 694

\bibitem[{{York} {et~al.}(2000){York}, {Adelman}, {Anderson}, {Anderson},
  {Annis}, {Bahcall}, {Bakken}, {Barkhouser}, {Bastian}, {Berman}, {Boroski},
  {Bracker}, {Briegel}, {Briggs}, {Brinkmann}, {Brunner}, {Burles}, {Carey},
  {Carr}, {Castander}, {Chen}, {Colestock}, {Connolly}, {Crocker}, {Csabai},
  {Czarapata}, {Davis}, {Doi}, {Dombeck}, {Eisenstein}, {Ellman}, {Elms},
  {Evans}, {Fan}, {Federwitz}, {Fiscelli}, {Friedman}, {Frieman}, {Fukugita},
  {Gillespie}, {Gunn}, {Gurbani}, {de Haas}, {Haldeman}, {Harris}, {Hayes},
  {Heckman}, {Hennessy}, {Hindsley}, {Holm}, {Holmgren}, {Huang}, {Hull},
  {Husby}, {Ichikawa}, {Ichikawa}, {Ivezi{\'c}}, {Kent}, {Kim}, {Kinney},
  {Klaene}, {Kleinman}, {Kleinman}, {Knapp}, {Korienek}, {Kron}, {Kunszt},
  {Lamb}, {Lee}, {Leger}, {Limmongkol}, {Lindenmeyer}, {Long}, {Loomis},
  {Loveday}, {Lucinio}, {Lupton}, {MacKinnon}, {Mannery}, {Mantsch}, {Margon},
  {McGehee}, {McKay}, {Meiksin}, {Merelli}, {Monet}, {Munn}, {Narayanan},
  {Nash}, {Neilsen}, {Neswold}, {Newberg}, {Nichol}, {Nicinski}, {Nonino},
  {Okada}, {Okamura}, {Ostriker}, {Owen}, {Pauls}, {Peoples}, {Peterson},
  {Petravick}, {Pier}, {Pope}, {Pordes}, {Prosapio}, {Rechenmacher}, {Quinn},
  {Richards}, {Richmond}, {Rivetta}, {Rockosi}, {Ruthmansdorfer}, {Sandford},
  {Schlegel}, {Schneider}, {Sekiguchi}, {Sergey}, {Shimasaku}, {Siegmund},
  {Smee}, {Smith}, {Snedden}, {Stone}, {Stoughton}, {Strauss}, {Stubbs},
  {SubbaRao}, {Szalay}, {Szapudi}, {Szokoly}, {Thakar}, {Tremonti}, {Tucker},
  {Uomoto}, {Vanden Berk}, {Vogeley}, {Waddell}, {Wang}, {Watanabe},
  {Weinberg}, {Yanny}, \& {Yasuda}}]{york00}
{York}, D.~G. {et~al.} 2000, \aj, 120, 1579

\bibitem[{{Zehavi} {et~al.}(1998){Zehavi}, {Riess}, {Kirshner}, \&
  {Dekel}}]{zehavi98}
{Zehavi}, I., {Riess}, A.~G., {Kirshner}, R.~P., \& {Dekel}, A. 1998, \apj,
  503, 483

\bibitem[{{Zheng} {et~al.}(2008){Zheng}, {Romani}, {Sako}, {Marriner},
  {Bassett}, {Becker}, {Choi}, {Cinabro}, {DeJongh}, {Depoy}, {Dilday}, {Doi},
  {Frieman}, {Garnavich}, {Hogan}, {Holtzman}, {Im}, {Jha}, {Kessler},
  {Konishi}, {Lampeitl}, {Marshall}, {McGinnis}, {Miknaitis}, {Nichol},
  {Prieto}, {Riess}, {Richmond}, {Schneider}, {Smith}, {Takanashi}, {Tokita},
  {van der Heyden}, {Yasuda}, {Assef}, {Barentine}, {Bender}, {Blandford},
  {Bremer}, {Brewington}, {Collins}, {Crotts}, {Dembicky}, {Eastman}, {Edge},
  {Elson}, {Eyler}, {Filippenko}, {Foley}, {Frank}, {Goobar}, {Harvanek},
  {Hopp}, {Ihara}, {Kahn}, {Ketzeback}, {Kleinman}, {Kollatschny},
  {Krzesi{\'n}ski}, {Leloudas}, {Long}, {Lucey}, {Malanushenko},
  {Malanushenko}, {McMillan}, {Morgan}, {Morokuma}, {Nitta}, {Ostman}, {Pan},
  {Romer}, {Saurage}, {Schlesinger}, {Snedden}, {Sollerman}, {Stritzinger},
  {Watson}, {Watters}, {Wheeler}, \& {York}}]{zheng08}
{Zheng}, C. {et~al.} 2008, \aj, 135, 1766

\bibitem[{{Zibin} {et~al.}(2008){Zibin}, {Moss}, \& {Scott}}]{zibin08}
{Zibin}, J.~P., {Moss}, A., \& {Scott}, D. 2008, Physical Review Letters, 101,
  251303

\end{thebibliography}

\end{document}